\documentclass[sageh]{sagej}
\usepackage{tikz}
\usetikzlibrary{shapes.geometric, arrows, matrix, positioning}
\usepackage{subcaption}
\usepackage{sectsty}
\usepackage{graphicx}
\usepackage{amsmath}
\usepackage{booktabs}
\usepackage{dsfont}
\usepackage{url}
\usepackage{amssymb}
\usepackage[draft]{changes}
\usepackage{comment}
\usepackage{bbm}
\usepackage[linewidth=1pt]{mdframed}
\usepackage{soul,color}
\usepackage{tikz}
\usepackage{setspace}
\usepackage{lineno}
\usetikzlibrary{decorations.pathreplacing}
\usetikzlibrary {arrows.meta} 
\def\ie{\textit{i.e.}}
\def\eg{\textit{e.g.}}

\title{A knapsack for collective decision-making}
\date{\today}

\author{Yurun Ge\affilnum{1,2}, Lucas B\"ottcher\affilnum{3,4}, Tom Chou\affilnum{2}, and Maria R.\ D'Orsogna\affilnum{1,2}}

\affiliation{\affilnum{1}Dept.~of Mathematics, California State University at Northridge, 91330 Los Angeles, USA\\
\affilnum{2}Dept.~of Computational Medicine, University of California, Los Angeles, 90095-1766  Los Angeles, USA\\
\affilnum{3}Dept.~of Computational Science and Philosophy, Frankfurt School of Finance and Management, Frankfurt am Main, Germany\\
\affilnum{4}Laboratory for Systems Medicine, Department of Medicine, University of Florida, 32611 Gainesville, FL, USA}

\corrauth{Maria R.\ D'Orsogna}

\email{dorsogna@csun.edu}

\begin{document}

\begin{abstract}
Collective decision-making is the process through which diverse
stakeholders reach a joint decision.  Within societal settings, one
example is participatory budgeting, where constituents decide on the
funding of public projects.  How to most efficiently aggregate diverse
stakeholder inputs on a portfolio of projects with uncertain long-term
benefits remains an open question.  We address this problem by
studying collective decision-making through the integration of
preference aggregation and knapsack allocation methods. Since different stakeholder groups may evaluate projects differently,
we examine several aggregation methods that combine their diverse
inputs.  The aggregated evaluations are then used to fill a
``collective'' knapsack.  Among the methods we consider are the
arithmetic mean, Borda-type rankings, and delegation to experts.  We
find that the factors improving an aggregation method’s ability to
identify projects with the greatest expected long-term value include
having many stakeholder groups, moderate variation in their expertise
levels, and some degree of delegation or bias favoring groups better
positioned to objectively assess the projects.  We also discuss how
evaluation errors and heterogeneous costs impact project selection.
Our proposed aggregation methods are relevant not only in the context
of funding public projects but also, more generally, for
organizational decision-making under uncertainty.
\end{abstract}


\maketitle
\section{Introduction}
The ethics and methods of collective decision-making have been the
subject of debate for centuries \citep{Blunden2016} and have a rich
history \citep{Bell1988,Schwenk1990,Schofield2002,Brandt2016,Baum2020}. Collective and fair decision-making is an
integral part of any functional democratic society
\citep{Lijphart1977,gersbach2005designing,gersbach2017redesigning}.  However, when making
decisions that affect diverse stakeholder groups, none of the various
alternatives may be perfectly ideal due to different priorities set by
the various groups, or cost limitations
\citep{Wahlstrom2001,Zahariadis2003}.  Decision-makers must thus
identify the best course of action compatible with various constraints
and the need for social cohesion. Furthermore, institutional
guardrails must guarantee transparency and confer legitimacy to the
process \citep{Glass1979}. As such, novel forms of community
engagement have emerged such as participatory design
\citep{hersh1999sustainable, Innes2004,gersbach2024forms}, where
stakeholders cooperate with urban planners or software designers on
land use or new technology \citep{Sanoff2006}, and participatory
budgeting, where decisions on how to spend public funds within a
strategic framework are delegated to the community \citep{Wampler2007,
  Aziz2021,benade2021preference,PBStanford2024}.

At the highest level is strategic decision-making made by governing
entities who set broad, long-term policies through economic, labor,
education, climate, or migration goals.  How a given vision is to be
implemented in practice, through the selection of actionable projects,
budgeting, partnerships, locations, define tactical decision-making of
medium-term impact.  Finally, operational decisions are ``on-the-ground'' choices that include personnel allocation and management,
timetables, or other logistic decisions that follow the strategic and
planning decisions and that are made accordingly.

Collective decision-making should
reflect, when possible, the majority stakeholder will, either by
direct voting or through delegation of agreed upon representatives.
Although one may identify optimal ways to solve social choice problems
within the confines of mathematical constructs, implementing them may
be accompanied with practical difficulties, or non-ideal conditions.
For example, ranked-voting methods, of which the Borda count is an
example, may be confusing for voters, leading to errors in ballot
marking, misunderstandings about how votes are aggregated, and voter
disenfranchisement.  Tallying votes may also be more complicated than
in traditional ``winner-takes-all'' voting, causing delays in
announcing results and necessitating sophisticated equipment and
personnel training.  It is illuminating that at the time of writing
(August 2024) several US states will be called to vote on
ranked-choice voting; some states aim to either repeal the method or
to prohibit it; others aim to adopt it.  Ten states have already
banned the practice.

In this paper, we focus on quantifying the efficiency of actionable choices in collective decision-making under cost constraints. We begin by analytically examining the properties of decision-making rules for a limited number of choices, before turning our attention to computational results. For concreteness, we embed our work in the context of
tactical decisions in the ``strategic-tactical-operational'' hierarchy
described above as it applies to societal decisions. For example, after setting the strategic goal of combating climate change, a jurisdiction may need to determine the appropriate mix of energy sources to exploit, decide whether to impose a carbon tax and on whom, choose between developing new technologies or enhancing existing ones, and evaluate whether to incentivize small-scale ``green'' energy projects or promote large-scale initiatives\textemdash and, if so, where to build the necessary infrastructure.

We will refer to each choice as a ``project''.  Several important
elements must be taken into account when modeling distinct parties
involved in collective decision-making.  The first is that although a
given ``project'' may have an intrinsic value, stakeholder groups will
evaluate it according to their own utility. Hence, there will be many
perceived values associated with the same project.  The second is that resources
are finite and that even a project that garners uniform consensus may
be discarded due to lack of funding or other forms of material
support.  Third, it is necessary to formulate clear and unequivocal
evaluation criteria so that the various alternatives can be ranked on
the basis of the preferences expressed by all groups, and the overall
most competitive projects be selected. Finally, while we primarily
focus on societal decision-making, the insights and methods presented
here are equally applicable to other settings where group decisions
are made~\citep{sah1988committees,Sah1986}, such as in corporations~\citep{csaszar2013organizational}, military
organizations~\citep{jaiswal2012military}, and medical diagnostics~\citep{Srivastava2022,Kurvers2023}.
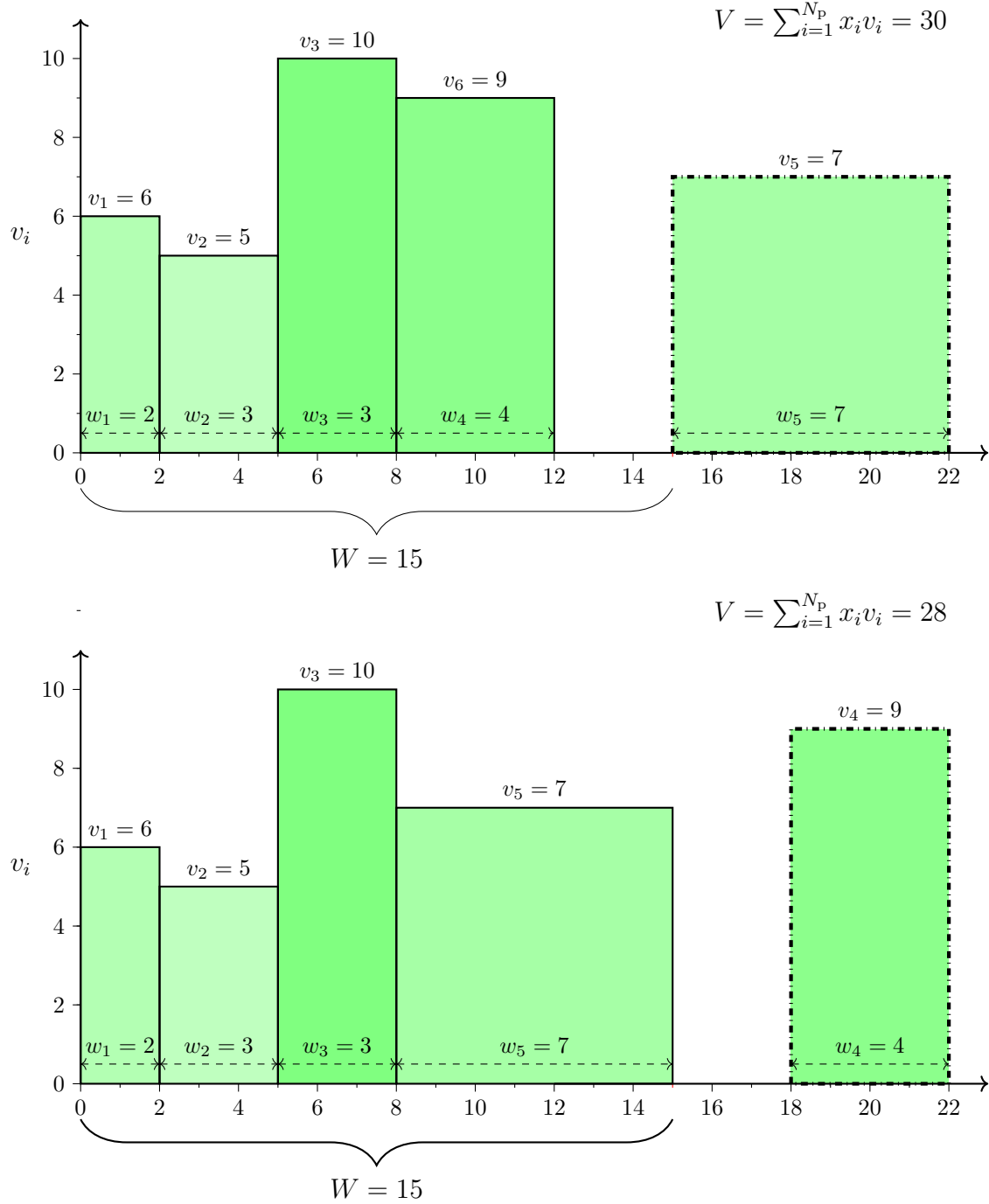
\begin{figure}[t!]
    \centering
    \begin{tikzpicture} [scale=0.6] 
    \fill[green!30] (0,0) rectangle (2,6);
    \fill[green!25] (2,0) rectangle (5,5);
    \fill[green!50] (5,0) rectangle (8,10);
    \fill[green!45] (8,0) rectangle (12,9);   
    \fill[green!35] (15,0) rectangle (22,7);
    \draw[dash dot, ultra thick] (15,0) rectangle (22,7);
     \node[above] at (18.5, 7) {$v_5=7$};
    \draw[thick,->] (12,0) -- (23,0) node[anchor=north] {};
    \draw[thick, ->] (0,0) -- (0,11) node[left, midway] {\large{$v_i$} \quad \,};    
\foreach \y in {0, 2, 4, 6, 8, 10} {
        \draw[](-0.2,\y) -- (0,\y) node[anchor=east] {$\y \, \, $};}
\foreach \y in {1, 3, 5, 7, 9} {
        \draw[](-0.1,\y) -- (0,\y); }
\foreach \x in {0, 2, 4, 6, 8, 10, 12, 14, 16, 18, 20, 22} {
        \draw[](\x, 0) -- (\x, -0.2) node[below] {$\x$};}
\foreach \x in {1, 3, 5, 7, 9, 11, 13, 17, 19, 21} {
        \draw[](\x, 0) -- (\x,-0.1); }
        \draw[red](15, 0) -- (15,-0.1); 
    \draw[thick] (0,0) rectangle (2,6) node[midway, above]{};
    \node[above] at (1,6) {$v_1=6$};
    \draw[thick] (2,0) rectangle (5,5) node[midway, above] {};
    \node[above] at (3.5, 5) {$v_2=5$};
    \draw[thick] (5,0) rectangle (8,10) node[midway, above] {};
    \node[above] at (6.5,10) {$v_3=10$};
    \draw[thick] (8,0) rectangle (12,9) node[midway, above] {};
    \node[above] at (10,9) {$v_6=9$};
    \draw[dashed, <->] (0,0.5) -- (2,0.5) node[midway, above] {$w_1=2$};
    \draw[dashed, <->] (2,0.5) -- (5, 0.5) node[midway, above] {$w_2=3$};
    \draw[dashed, <->] (5,0.5) -- (8,0.5) node[midway, above] {$w_3=3$};
    \draw[dashed, <->] (8,0.5) -- (12,0.5) node[midway, above] {$w_4=4$};
    \draw[dashed, <->] (15,0.5) -- (22,0.5) node[midway, above] {$w_5=7$};
    \draw[black, decorate,decoration={brace, amplitude=20pt, mirror}] (0,-0.9) -- (15,-0.9) 
    node[black, midway, yshift=-30pt] {\large{\textbf{$W=15$}}};
                  \node at (19, 11) {\large{\textbf{$V=\sum_{i=1}^{N_{\rm p}} x_i v_i = 30$}}};
    \fill[green!30] (0,-16) rectangle (2,-10);
    \fill[green!25] (2,-16) rectangle (5,-11);
    \fill[green!50] (5,-16) rectangle (8,-6);
    \fill[green!35] (8,-16) rectangle (15,-9);   
    \fill[green!45] (18,-16) rectangle (22,-7);
    \draw[dash dot, ultra thick] (18,-16) rectangle (22,-7);
         \node[above] at (20, -7) {$v_4=9$};
    \draw[thick,->] (12,-16) -- (23,-16) node[anchor=north] {};
    \draw[thick, ->] (0,-16) -- (0,-5) node[midway, left] {\large{$v_i$} \quad \,};    
\foreach \y/\i in {-16/0, -14/2, -12/4, -10/6,  -8/8, -6/10} {
        \draw[](-0.2,\y) -- (0,\y) node[anchor=east] {$\i \, \, $};}
\foreach \y in {-14, -12, -10,  -8,  -6, -4} {
        \draw[](-0.1,\y) -- (0,\y); }
\foreach \x in {0, 2, 4, 6, 8, 10, 12, 14, 16, 18, 20, 22} {
        \draw[](\x, -16) -- (\x, -16.2) node[below] {$\x$};}
\foreach \x in {1, 3, 5, 7, 9, 11, 13, 17, 19, 21} {
        \draw[](\x, -16) -- (\x,-16.1); }
        \draw[red](15, -16) -- (15,-16.1); 
    \draw[thick] (0,-16) rectangle (2,-10) node[midway, above]{};
    \node[above] at (1,-10) {$v_1=6$};
    \draw[thick] (2,-16) rectangle (5,-11) node[midway, above] {};
    \node[above] at (3.5, -11) {$v_2=5$};
    \draw[thick] (5,-16) rectangle (8,-6) node[midway, above] {};
    \node[above] at (6.5,-6) {$v_3=10$};
    \draw[thick] (8,-16) rectangle (15,-9) node[midway, above] {};
    \node[above] at (11.5,-9) {$v_5=7$};
    \draw[dashed, <->] (0,-15.5) -- (2,-15.5) node[midway, above] {$w_1=2$};
    \draw[dashed, <->] (2, -15.5) -- (5, -15.5) node[midway, above] {$w_2=3$};
    \draw[dashed, <->] (5, -15.5) -- (8,-15.5) node[midway, above] {$w_3=3$};
    \draw[dashed, <->] (8,-15.5) -- (15,-15.5) node[midway, above] {$w_5=7$};
    \draw[dashed, <->] (18,-15.5) -- (22,-15.5) node[midway, above] {$w_4=4$};
    \draw[black, thick, decorate,decoration={brace, amplitude=20pt, mirror}] (0,-16.9) -- (15,-16.9) 
    node[black, midway, yshift=-30pt] {\large{\textbf{$W=15$}}};
                  \node at (19, -4) {\large{\textbf{$V=\sum_{i=1}^{N_{\rm p}} x_i v_i = 28$}}};
\end{tikzpicture}
    \captionsetup{justification=justified}
    \caption{Schematic of filling a knapsack of maximal weight $W$
      with a subset of a group of items carrying weights $w_{i}$ and
      values $v_{i}$ for $i \in \{1, \dots, N_{\rm p} \}$.  Here,
      $W=15$ and $N_{\rm p} = 5$.  The weights $w_{i}$ are represented
      by the width of each item while their values $v_{i}$ are
      indicated by their heights and by the green intensity.  Panel
      (a) depicts the optimal selection which requires rejection of
      item $i=5$ so that $x_i =1, i \in \{1, \dots, 4 \}$, and $x_5
      =0$ in Eq.\,\eqref{max}.  This selection yields a total knapsack
      value $V = \sum_i x_i v_i = 30$ but leaves three units of
      capacity unoccupied since the total weight is $\sum_i x_i w_i =
      12$.  The rejected item $i=5$ is depicted as a dot-dashed
      rectangle.  Panel (b) shows an alternative way of filling the
      knapsack with items that reach the knapsack weight capacity $W
      =15$. Here, the rejected item is $i=4$.  However, this knapsack
      is suboptimal because its total value is $V=\sum_i x_i v_i = 28
      < 30$. The quality of each item is represented by each
      rectangle's aspect ratio $q_{i} = v_{i}/w_{i}$ for $i \in \{1,
      \dots, N_{\rm p} \}$.
    }
    \label{fig:knapsack_demo}
\end{figure}
\section{Knapsack problems}
Our analysis frames social choice in the context of the so-called
``knapsack problem'', which was first formulated in the late 1940s. It refers to the problem of selecting a set of items, each with a given weight and value, to maximize their total value without exceeding the knapsack's weight limit. The binary knapsack problem is mathematically framed as an ``extremum problem'', where the goal is to find the optimal solution to a linear system~\citep{dantzig1957discrete}.  More formally, given
a set of $N_{\rm p}$ items (or projects), each characterized by value
$v_i > 0 $ and weight $w_i > 0$ for ${i\in\{1,\dots, N_{\rm p}\}}$,
the binary knapsack problem consists of finding $x_i \in \{0,1\}$ for
each $i$ such that
\begin{align}
 &   \sum_{i=1}^{N_{\rm p}} x_i v_i   \quad \mbox{is maximal, subject to the constraint}
    \label{max} \\
&    \sum_{i=1}^{N_{\rm p}} x_i w_i \leq W,  \quad \mbox{with $x_i \in \{0,1 \}$}, 
    \label{constraint}
\end{align}
where $W$ is the total weight supported by the knapsack and where the
indicator $x_i\in\{0,1\}$ represents whether item $i$ has been
selected ($x_i=1$) or not ($x_i=0$) \citep{Pisinger1998}.  A schematic
is given in Fig.\,\ref{fig:knapsack_demo}.  Variants of this problem
include the bounded knapsack where $x_i \in \{0,1, \dots c\}$ so that
a maximum of $c \geq 1$ copies of each item can be selected, with $c
\to \infty$ in the unbounded version of the problem.  Similarly, many
methods have been proposed to obtain optimal solutions, including
dynamic programming \citep{martello1987algorithms}, where the weights
$w_i $ and the total weight $W$ are integers, and branch-and-bound
algorithms~\citep{dudzinski1987exact,Pisinger1998} that work well for
a limited number of items. As computing capabilities increased over time,
and as several knapsack problems were cleverly solved, more
challenging constraints were added to make solutions more difficult,
for example imposing correlations between weights and values of the
knapsack items.  The search for ``hard knapsacks" continues to this
day \citep{nasako2006, pisinger2005hard, smith2021revisiting,
  Cacchiani2022}.

Within social choice theory, knapsack problems may serve as paradigms
for collective decision-making under constraints.  For example, the
set of $N_{\rm p}$ items to be included in the knapsack can represent
possible public projects or scheduling options presented to the
community; the value $v_i$ of each item $i$ to be included in the
knapsack can be interpreted as the project's utility or benefit; the
weight $w_i$ can represent monetary or other forms of cost.  The
knapsack itself can be viewed as the final portfolio of chosen
projects, the maximal weight $W$ representing the total budget.  But
who is filling the knapsack?  As discussed above, different
stakeholder groups will partake in the same public decision-making
process and each of them will assign subjective values to the same
projects.  Thus, each group may have their own $v_i, x_i$ values and
may fill the knapsack in a way that is unique to their own experience.
The goal then is to optimally aggregate the multiple preferences under
the budgetary constraint $W$ to fill a ``community-knapsack" born of
all the various inputs.  This is where preference aggregation methods
and voting rules interface with knapsack-like problems.  For example,
community preferences expressed in participatory budgeting can be
aggregated using ad-hoc knapsack schemes that include actual revenues,
deficits, or surpluses \citep{Goel2019}.  Some existing algorithms to
fill the common knapsack aim to prioritize total value regardless of
community preference, or the diversity or fairness of the final
knapsack selection \citep{Fluschnik2019}.  Other approaches include
the use of cooperative game theory or determining the Shapley value so
that once a common knapsack is filled, resulting benefits can be
shared equitably among participant stakeholders
\citep{Arribillaga2022, Bhagat2014}.  Fair division and cake-cutting
problems on the other hand involve multiple agents dividing or
consolidating a set of assets as in the case of inheritance, divorces,
or corporate mergers. These can be studied via divisible knapsacks
where each agent can fill their own knapsack with fractional parts of
common items to ensure fair allocation of resources \citep{Brams1996}.

\section{Collective decisions to fill the knapsack}
In this paper, we focus on collective decision-making in the context
of tactical choices; one may use the sustainability and environmental
sphere as a concrete setting.  We assume there are a set of $N_{\rm
  p}$ projects (or measures) presented to the community; these are the
items to select from in the knapsack problem described above. The
societal long-term benefit that derives from project $i$ is given by
$v_i$ for $i \in \{1, \dots, N_{\rm p} \}$.  For example, $v_i$ may
represent how much measure $i$ reduces greenhouse gas emissions or how
much renewable energy the construction of wind farm $i$ will yield
over a specific time frame.  Similarly, each potential project is
associated with a cost $w_i$.  We assume $v_i >0 $ and $w_i >0$ to be
fixed and to be objectively measurable, so that a single
decision-making entity with perfect knowledge of the $v_i, w_i$ values
would be left with the task of selecting a subset of projects with
highest value under the given budget $W$. Mathematically, the problem
is that of solving the binary knapsack stated in Eqs.\,\eqref{max} and
\eqref{constraint}.

\begin{table}[t!]
\centering
\begin{tabular}{c l}
\toprule
Symbol & Description \\ \midrule
$W > 0$
& Total available budget  \\
$N_{\rm p} \in\mathbb{Z}^{+}$ & Number of projects \\
$N_{\rm s}\in\mathbb{Z}^{+}$ & Number of stakeholder groups \\
$i \in \{1, \dots, N_{\rm p} \}$  & Project label \\
$j \in \{1, \dots, N_{\rm s} \}$ & Group label  \\
$v_i > 0$
& Value of project $i$ \\ 
$w_i > 0$
& Cost of project $i$ \\
$q_i >0 $
& Quality of project $i$; \, $q_i = v_i / w_i$ \\
$t_i\in[t_{\rm min}, t_{\rm max}]$ & Type of project $i$  \\
$e_j \in[e_{\rm min}, e_{\rm max}]$ 
& Expertise of group $j$  \\
$t_{\rm M}$
& Average project type; $t_{\rm M} = (t_{\rm max} + t_{\rm min})/2$ \\
$e_{\rm M}$
& Average expertise level; $e_{\rm M} = (e_{\rm max} + e_{\rm min})/2$ \\
$\beta \geq 0 $ & Knowledge breadth of groups;  $ \beta =  (e_{\rm max} - e_{\rm min})/2$ \\
$\sigma_{ij} \geq 0 $
& Perception error when project $i$ is evaluated by group $j$\\
$v_{ij}$
& Value of project $i$ as evaluated by group $j$ \\
$q_{ij}$
& Quality of project $i$ as evaluated by group $j$; $q_{ij}=v_{ij}/w_i$ \\
$v'_i$
&  Aggregate value of project $i$ over all $N_{\rm s}$ groups \\
$q'_{i}$
& Aggregate score of project $i$  \\
\bottomrule
\end{tabular}
\captionsetup{justification=justified}
\vspace{0.2mm}
\caption{An overview of the main model parameters and variables.  
Unless specified, all parameters can take real values. 
Although $e_{\rm M}, t_{\rm M}$ can be independent, we set $e_{\rm M} = t_{\rm M}$  to signify that the central expertise level is also the central project type. 
The main variables in this work are $N_{\rm s}, N_{\rm p}, \beta$ whereas we fix
$ e_{\rm M} = t_{\rm M} =5,  t_{\rm min} =0, t_{\rm max} =10$,  
unless stated otherwise. For mathematical convenience and without loss of generality we choose $N_{\rm p}$ to be even
and $N_{\rm s}$ to be odd.  We consider
two scenarios: one where costs are uniform, $w_i = 1$, and another where 
they define a decreasing function of $i$. }
\label{tab:model_quantities}
\end{table}

However, instead of a single entity, we assume there is a diverse set of $N_{\rm s}$ stakeholder groups 
that are affected by the proposed projects and who partake in the decision-making process.
We assume that  each of the $j \in \{1, \dots, N_{\rm s} \}$ groups assigns 
a different value to project $i$, on the basis of their specific experience, expertise or priority.  
We thus define the perceived value $v_{ij}$  as the value of project $i$ as per the prerogatives
of group $j$. The $v_{ij}$ evaluations may include both positive and negative contributions to the objective
value $v_i$ so that for some $j$ groups $v_{ij}  >v_i$, for others $v_{ij} < v_i$, for others yet $ v_{ij} =v_i$. 
For example, the overall benefit $v_i$ of establishing the large-scale renewable facility $i$  may be lessened 
by the necessary perturbation or even destruction of historical or sacred sites to accommodate the facility or related
infrastructure, or by the loss of tourism or agriculture. These may carry special cultural or economic 
relevance for group $j$, for which $v_{ij} < v_i$. In other scenarios, the same facility would
be accompanied with new job opportunities for members of group $k$; in this case $v_{ik} > v_i$. 
We keep $w_i$, the cost of project $i$, independent of groups $j,k$. 
All variables are listed in Table\,\ref{tab:model_quantities}. For
simplicity, and without loss of generality we choose $N_{\rm p}$ to be even
and $N_{\rm s}$ to be odd. 

The main question we address in this paper is this: given that each group $j$ has its own
set of preferences for the $N_{\rm p}$ projects, 
how does a central decision-maker or agency decide which projects to select for final approval, 
assuming there is a total budget $W$? 
One way to do this is to construct a ``community-knapsack" and solve
Eqs.\,\eqref{max} and \eqref{constraint} using the given costs $w_i$ and introducing 
an effective value $v_i'$ for each item $i$ that is derived from the group evaluations $v_{ij}$. 
To determine $v_i'$ the central decision-maker or agency must have an understanding 
of how the various perceived values $v_{ij}$ are determined and the degree of heterogeneity among groups
so that appropriate methods can be employed. 
Although a natural approach would be to 
equate $v_i'$ with the arithmetic mean across the $v_{ij}$ evaluations given by 
the $N_{\rm s}$ groups, there are many other possible aggregation methods. We examine them in the next 
section building on previous work on optimal portfolio selection where an organization 
(the equivalent of our central decision-maker or agency)
must select which future investments to pursue (the equivalent of selecting the items that belong in the knapsack)
taking input from various agents (the equivalent of our community groups) whose evaluation of
the value of projects is characterized by so-called perceptual noise 
\citep{csaszar2013organizational, boettcher2024selection}. 
In this existing body of work, concepts such as group expertise, knowledge breadth, and project type are included in the 
decision-making process, and we will adapt them to our specific scenario. 
The model proposed by \cite{boettcher2024selection} also focuses on portfolio selection of projects. 
However, it does not incorporate heterogeneous costs $w_i$, 
an assumption we will here relax. 

\subsection{Portfolio selection by each group}
In this section, we describe how the $N_{\rm s}$ groups determine their $v_{ij}$ evaluations
of project $i$.  For simplicity, we set $v_{ij} = v_i + \epsilon_{ij}$ and discuss various choices
for $\epsilon_{ij}$, which may be positive or negative. 
One possible scenario is that groups may be able to accurately assess 
tangible detrimental, beneficial or neutral effects of the project; in this case there is no uncertainty in
determining $v_{ij}$ and we may set $\epsilon_{ij} = \epsilon^{\rm t}_{ij}$. 
In other situations,  however, groups may not have the in-house expertise for a proper evaluation of the project
and $v_{ij}$ may include perception uncertainties, rather than estimations of
actual detriment or benefits to the community. In this case, we set $\epsilon_{ij} = \epsilon^{\rm p}_{ij}$.
Although the two contexts are distinct, both of them 
lead to subjective assessments $\epsilon_{ij}$
that we add (or subtract) to the underlying value $v_i$. 

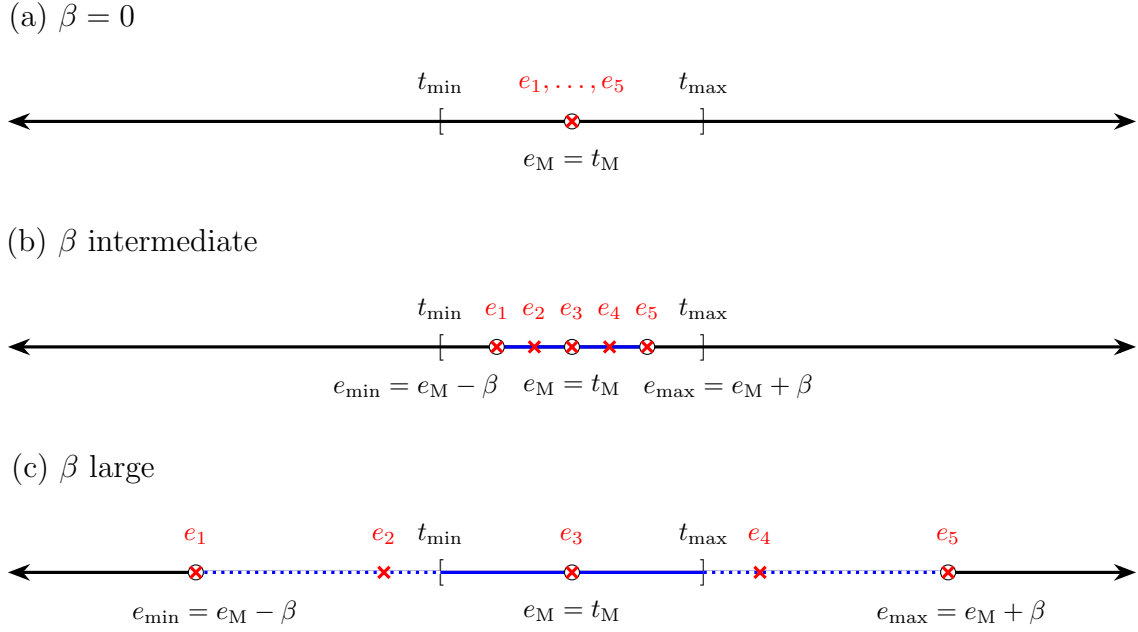
\begin{figure}[t]
    \centering
\begin{tikzpicture} [scale=2.5] 
	\tikzset{
	mycross/.pic={ \draw[red, very thick, rotate=45, pic actions] 
		               (-3pt,0) -- (3pt,0)
	        	        	      (0,-3pt) -- (0,3pt);
	        		       }}
		\draw [->,  very thick, Stealth-Stealth] (-3, 1.25) -- (3, 1.25);   
   	\node[black] at (-0.7, 1.25) {[}; 
      	\node[black] at (0.7, 1.25) {]}; 
    	\node[circle, draw=black, fill= white, inner sep=2pt, minimum size=5pt] at (0, 1.25){};
    	 \node[font=\normalfont, align=center, below] at (0, 1.15) {$e_{\rm M} = t_{\rm M}$};
       	\node[font=\normalfont, align=center, above] at (-0.7, 1.35) {$t_{\rm min} $}; 
	             \node[font=\normalfont, align=center, above] at (0.7, 1.35) {$t_{\rm max}$};
	            \foreach \x in {0}
	            { 
	             \pic at (\x, 1.25) {mycross}; }  
	             \node[font=\normalfont, red, align=center, above] at (0, 1.35) {$e_{1}, \dots ,e_5$}; 
				     									 \node[align=left] at (-2.66, 1.80)  {\large {(a) $\beta=0$}};
	                \draw [->,  very thick, Stealth-Stealth] (-3, 0.05) -- (3, 0.05);   
   \node[black] at (-0.7, 0.05) {[}; 
      \node[black] at (0.7, 0.05) {]}; 
           \draw[blue, very thick] (-0.4, 0.05) -- (0.4, 0.05){};
      \node[font=\normalfont, align=center, below] at (-0.83, -0.05)  {$e_{\rm min} = e_{\rm M} - \beta$};
      \node[font=\normalfont, align=center, below] at (0.83, -0.05) {$e_{\rm max} = e_{\rm M} + \beta$};
    \node[circle, draw=black, fill= white, inner sep=2pt, minimum size=5pt] at (0, 0.05){};
    	 \node[font=\normalfont, align=center, below] at (0, -0.05) {$e_{\rm M} = t_{\rm M}$};
	 	 \node at (-0.4,0.05) [circle, draw=black, fill= white, inner sep=2.pt]{};
       		\node[font=\normalfont, align=center, above] at (-0.7, 0.15) {$t_{\rm min} $}; 
		             	\node at (0.4,0.05) [circle, draw=black, fill= white, inner sep=2.pt]{};
	             \node[font=\normalfont, align=center, above] at (0.7, 0.15) {$t_{\rm max}$};
	            \foreach \x in {-0.4,-0.2 ,0, 0.2, 0.4}
	            { 
	             \pic at (\x,0.05) {mycross}; }  
	             \node[font=\normalfont, red, align=center, above] at (-0.4, 0.15) {$e_{1}$}; 
				       		\node[font=\normalfont, red, align=center, above] at (-0.2, 0.15) {$e_{2}$}; 
						       		\node[font=\normalfont, red, align=center, above] at (0, 0.15) {$e_{3}$}; 
								       		\node[font=\normalfont, red, align=center, above] at (0.2, 0.15) {$e_{4}$}; 
										       		\node[font=\normalfont, red, align=center, above] at (0.4, 0.15) {$e_{5}$};  
												 \node[align=right] at (-2.34, 0.6)  {\large {(b) $\beta$ intermediate}};
 \draw [->,  very thick, Stealth-Stealth] (-3, -1.15) -- (3, -1.15);    
   \node[align=left] at (-2.6, -0.6)  {\large {(c) $\beta$ large}};
   \node[black] at (-0.7, -1.15) {[}; 
      \node[black] at (0.7, -1.15) {]}; 
      \node[font=\normalfont, align=center, above] at (-0.7, -1.05) {$ t_{\rm min} $};
      \node[font=\normalfont, align=center, above] at (0.7, -1.05) {$t_{\rm max}$};
       \draw[blue, very thick] (-2, -1.15) -- (2, -1.15){};
             \draw[white, very thick] (-2, -1.15) -- (-0.7, -1.15){};
                       \draw[blue, dotted, very thick] (-2, -1.15) -- (-0.7, -1.15){};
                          \draw[white, very thick] (0.7, -1.15) -- (2, -1.15){};
                               \draw[blue, dotted, very thick] (0.7, -1.15) -- (2, -1.15){};
    \node[circle, draw=black, fill= white, inner sep=2pt, minimum size=5pt] at (0, -1.15){};
    	 \node[font=\normalfont, align=center, below] at (0, -1.25) {$e_{\rm M} = t_{\rm M}$};
	 	 \node at (-2, -1.15) [circle, draw=black, fill= white, inner sep=2.pt]{};
       		\node[font=\normalfont, align=center, below] at (-1.9, -1.25) {$e_{\rm min} = e_{\rm M} - \beta$};
        	\node at (2,-1.15) [circle, draw=black, fill= white, inner sep=2.pt]{};
	             \node[font=\normalfont, align=center, below] at (2.07, -1.25) {$e_{\rm max} = e_{\rm M} + \beta$};
	             	            \foreach \x in {-2,-1 ,0, 1, 2}
	            { 
	             \pic at (\x,-1.15) {mycross}; }  
	             \node[font=\normalfont, red, align=center, above] at (-2, -1.05) {$e_{1}$}; 
				       		\node[font=\normalfont, red, align=center, above] at (-1, -1.05) {$e_{2}$}; 
						       		\node[font=\normalfont, red, align=center, above] at (0, -1.05) {$e_{3}$}; 
								       		\node[font=\normalfont, red, align=center, above] at (1, -1.05) {$e_{4}$}; 
										       		\node[font=\normalfont, red, align=center, above] at (2, -1.05) {$e_{5}$};   		   	                   
												   \end{tikzpicture}
     \captionsetup{justification=justified}
\vspace{0.2mm}
     \caption{Schematic of possible distributions of the project types
       $t_i \in [t_{\rm min}, t_{\rm max}]$ and expertise levels $e_i
       \in [e_{\rm min}, e_{\rm max}]$ for $N_{\rm s} = 5$ projects
       whose expertise levels are distributed according to
       Eq.\,\eqref{expert}.  The respective midpoints are given by
       $t_{\rm M} = (t_{\rm max} + t_{\rm min})/2 $ and $e_{\rm M} =
       (e_{\rm max} + e_{\rm min})/2 $.  In this paper, we assume the
       two coincide, so that $t_{\rm M} = e_{\rm M}$.  We use $\beta$
       to denote the spread of the expertise level, $\beta =( e_{\rm
         max} - e_{\rm min})/ 2$.  By construction, $e_1 = e_{\rm
         min}$, $e_3 = e_{\rm M}$, and $e_5 = e_{\rm max}$ in all
       panels.  In panel (a), we set $\beta =0$ so that all groups
       have the same expertise, $e_j = e_{\rm M} = t_{\rm M}$ for all
       $j \in \{1, \dots, N_{\rm s} \}$.  The lack of diversity may
       hinder the proper evaluation of projects $i$ whose type $t_i$
       is different from $t_{\rm M}$.  Thus, some of the resulting
       perception errors $\sigma_{ij} = |t_i - e_j|$ may be relatively
       large.  In panel (b), we choose an intermediate value of
       $\beta$ so that for any project $i$ of type $t_i$ there may be
       one or more groups $j$ that are well positioned to evaluate it.
       All resulting perception errors $\sigma_{ij} = |t_i - e_j |$
       will be relatively small.  In panel (c), $\beta$ is largest.
       Here, the expertise spread exceeds that of the $[t_{\rm min},
         t_{\rm max}]$ interval so that some groups $j$ may not have
       the expertise to evaluate any project type $i$. For these
       groups the perception errors $\sigma_{ij} = |t_i - e_j|$ will
       be relatively large. This is the case for group $j=1$ whose
       expertise level $e_1 = e_{\rm min} \ll t_{\rm min}$ and for
       group $5$ whose expertise level $e_5 = e_{\rm max} \gg t_{\rm
         max}$.  In our numerical examples we set $t_{\rm min} =0,
       t_{\rm max} =10$; $t_{\rm M} = e_{\rm M} =5$ and vary $N_{\rm
         s}$ and $\beta$.  }
         \label{fig:knowledge_breadth_demo}
\end{figure}

If communities have a clear perception of the consequences of a
project, one way to model $\epsilon^{\rm t}_{ij}$ is to assume that
benefits ($+ \delta_i$) and disadvantages ($- \delta_i)$ represent a
zero-sum game so that $\epsilon^{\rm t}_{ij} = \chi_j \delta_i$ where
$\chi_j = \{-1,0,1 \}$ and $\sum_{j} \chi_j \delta_i = 0$.  Non
zero-sum scenarios would restrict $\chi_j$ to the set $\chi_j = \{0,1
\}$, where project $i$ is neutral or beneficial to communities, or to
the set $\chi_j = \{-1,0\}$, where project $i$ is neutral or
detrimental.  Another possibility is to assume that $\epsilon^{\rm
  t}_{ij}$ is a random variable chosen from a normal distribution with
zero mean and standard deviation $\sigma_{ij} = \sigma$ so that
$\epsilon^{\rm t}_{ij}\sim\mathcal{N}(0,\sigma)$.  This choice implies
that the nature and consequence of each project is well understood by
all groups, as described above, but the impacts can be quite different
across them. Large values of $\sigma$ will result in large
heterogeneities in $v_{ij}$, representing diverse communities who will
judge project $i$ along a large spectrum of actual benefits and
disadvantages; small values of $\sigma$ represent more aligned
communities whose evaluation spread is more limited.  Zero-sum is not
guaranteed in this setting.

If communities lack the proper tools to analyze or forecast the
impacts of a project, we can still pose $\epsilon^{\rm
  p}_{ij}\sim\mathcal{N}(0,\sigma_{ij})$ but describe $\sigma_{ij}$
through concepts such as type $t_i$ of project $i$, and expertise
$e_j$ of group $j$, following the perceptual noise description of
\citep{csaszar2013organizational,boettcher2024selection}.  The
procedure to determine the standard deviation $\sigma_{ij}$ associated
with the perception noise is as follows. Two projects $i$ and $i'$ are
assumed to be of the same type ($t_i = t_{i'}$) if they fall within
comparable categories (\eg, both projects are focused on reforestation
and wildfire prevention, or both projects are focused on renewable
energy generation). We assume that $t_i$ is taken from a uniform
distribution $\mathcal{U}(t_{\rm min},t_{\rm max})$, with $t_{\rm min}
< t_{\rm max}$. Analogously, two groups $j$ and $j'$ have the same
expertise ($e_j=e_{j'}$) if their know-hows are closely aligned.  For
concreteness, we further formalize the definition of the levels of
expertise $e_j$ by setting
\begin{equation}
\label{expert}
	e_j = e_{\rm M} - \frac{N_{\rm s} + 1 - 2 j}{N_{\rm s}-1} \beta, \qquad {j \in \{1, \dots, N_{\rm s} \}}.
\end{equation}
Equation\,\eqref{expert} implies that the $e_j$ values are equally
spaced on the interval $[e_{\rm M} - \beta, e_{\rm M} + \beta] \equiv
[e_{\rm min}, e_{\rm max}]$; $e_{\rm M}$ is the average expertise
level and $\beta$ is the knowledge breadth that determines their
spread.  The magnitudes $t_i, e_j$ carry no specific significance,
they are just labels to represent different types and expertises.

We now set $\sigma_{ij} \equiv |t_i - e_j|$ which implies that if the
type of project $i$ matches exactly the expertise of group $j$ and
$t_i = e_j$, then $\sigma_{ij} = 0$.  In this case, there are no
errors in the evaluation of project $i$ by group $j$, and $v_{ij} =
v_i$. On the other hand, a large discrepancy between project type and
group expertise leads to a large variability in the possible
evaluation of project $i$ which can be designated to be either
unrealistically beneficial or unrealistically detrimental. This would
be encapsulated by a large $\sigma_{ij} = |t_i - e_j|$.  The modeling
choices for $\sigma_{ij}$, types $t_i$ and expertise levels $e_j$ are
occasionally referred to as the Hotelling-type
model~\citep{csaszar2013organizational,boettcher2024selection}, named
after Harold Hotelling who studied competition among shops that are
located on a street or products whose properties are distributed along
an interval~\citep{hotelling1929stability,NOVSHEK1982199}.  We will
refer to $\sigma_{ij}$ as the perception error.  A schematic for
$N_{\rm s} = 5$ groups whose expertise levels are distributed
according to Eq.\,\eqref{expert} for three representative values of
$\beta$ is shown in Fig.\,\ref{fig:knowledge_breadth_demo}.

A more realistic representation of $v_{ij}$ would incorporate both
tangible benefits or disadvantages and perception noise leading to
$v_{ij} = v_i + \epsilon^{\rm t}_{ij} + \epsilon^{\rm p}_{ij}$ with
$\epsilon^{\rm t}_{ij} = \chi_{j} \delta_i$ or $\epsilon^{\rm t}_{ij}
\sim {\cal N} (0, \sigma)$ representing the specific additional impact
of project $i$ on group $j$, and $\epsilon^{\rm p}_{ij} \sim {\cal N}
(0, \sigma_{ij})$ representing the noise associated to group $j$
evaluating project $i$.  For simplicity we only consider normally
distributed perception noise $v_{ij} = v_i + \epsilon^{\rm p}_{ij}$
with $\epsilon^{\rm p}_{ij} \sim {\cal N} (0, \sigma_{ij})$ and
perception error $\sigma_{ij} = |t_i - e_j|$.

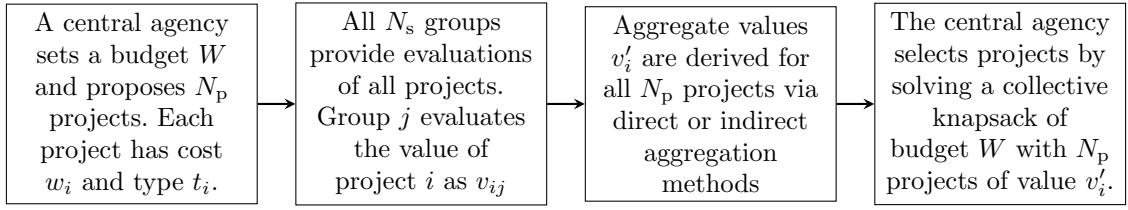
\begin{figure}[t!]
\centering 
\begin{tikzpicture}[
    node distance=1cm and 0.5cm, 
    startstop/.style={rectangle, rounded corners, text width=3cm, minimum height=1cm, text centered, draw=black, 
    align=center},
    io/.style={rectangle, text width=3.1cm, minimum height=1cm, text centered, draw=black, 
    align=center},
    process/.style={rectangle, text width=3.1cm, minimum height=1cm, text centered, draw=black, 
    align=center},
    arrow/.style={thick,->,>=stealth}
]
\node (input) [io] {A central agency sets a budget $W$ and proposes $N_{\rm p}$ projects. Each project has cost $w_i$ and type $t_i$. };
\node (eval) [process, right=of input] {All $N_{\rm s}$ groups provide evaluations of all projects.  Group $j$ evaluates the value of project $i$ as $v_{ij}$};
\node (aggregate) [process, right=of eval] {Aggregate values $v'_i$ are derived for all $N_{\rm p}$ projects via direct or indirect aggregation methods};
\node (decision) [io, right=of aggregate] {The central agency selects projects by solving a collective knapsack of budget $W$ with $N_{\rm p}$ projects of value
$v_i'$.};

\draw [arrow] (input) -- (eval);
\draw [arrow] (eval) -- (aggregate);
\draw [arrow] (aggregate) -- (decision);

\end{tikzpicture}
    \captionsetup{justification=justified}
\caption{A flowchart illustrating the collective decision-making
  process under cost constraints. The central agency or elected
  decision-maker sets a strategic goal and allocates a budget
  $W$. Stakeholders are presented with $N_{\rm p}$ actionable projects
  and form $N_{\rm s}$ distinct groups whose priorities and views may
  align or be in opposition. }
\label{fig:flowchart} 
\end{figure}

Once all estimates $v_{ij}$ are determined, one must aggregate them in
order to determine the effective values $v_i'$ and fill the community
knapsack through Eqs.\,\eqref{max} and \eqref{constraint}, as shown in
Fig.\,\ref{fig:flowchart}. We assume that each group $j$ ``submits"
their own list of values $v_{ij}$ in ascending order ($v_{(1) j} \leq
v_{(2) j} \leq \dots \leq v_{(N_{\rm p}) j}$ for all $j$) so that the
central decision-maker or agency can aggregate them using one of the
alternatives described in the next section.

\section{Aggregation methods}

We now introduce various methods for aggregating the evaluation lists
$ \{v_{ij} \}$ submitted by each of the $N_{\rm s}$ groups in
preparation for the final decision-making.  These methods can be
broadly categorized into direct and indirect approaches. Direct
approaches aim to synthesize the individual estimates $v_{ij}$ for $j
\in \{1, \dots, N_{\rm s} \}$ into an effective $v'_{i}$, for example
through the use of the arithmetic mean. Indirect approaches are based
on value proxies such as relative rankings among projects; a project
with a majority of favorable votes is considered superior to others.
Among the indirect methods are ones where we rescale evaluations and
project them onto standard ranges.  We overview the two methods in the
next subsections under perception noise, $\epsilon_{ij} =
\epsilon^{\rm p}_{ij} \sim {\cal N} (0, \sigma_{ij})$, with perception
noise $\sigma_{ij} = |t_i - e_j|$.

\subsection{Direct aggregation methods}
    \label{SSAM}

Here, we list various aggregation methods that yield an effective
value $v_i'$ for each item $i$ based on the inputs $v_{ij} $ from all
$N_{\rm s}$ groups. In some cases, the variances $\sigma_{ij}$ are
used, in others they are not.  Once the values $v_i'$ are determined
the collective knapsack is filled by optimizing the following problem

\begin{equation}
    \max \sum_{i=1}^{N_{\rm p}} x_i v'_i
    \label{max3}
\end{equation}
subject to the constraint in Eq.\,\eqref{constraint}. Note that the
collective knapsack is filled via the aggregated values $v_i'$ that
may, or may not, be close to the intrinsic $v_i$ values.

\subsubsection{Arithmetic mean: }
The arithmetic mean is the most common estimator of central tendency
for random variables.  Here, we use it as an aggregation
method. Assuming that all group inputs are given equal consideration,
$v'_{i}$ is given by

\begin{equation}
 v'_i = \frac 1 {N_{\rm s}} \sum_{j=1}^{N_{\rm s}}v_{ij}. 
\end{equation}

\subsubsection{Asymmetric estimators and the Median:}
 
Other useful aggregation methods are based on ordering statistics
\citep{stuart2010kendall} whereby the $v_{ij}$ evaluations for project
$i$ are appropriately sorted and the input of some groups are
prioritized over those of others. For the sake of discussion, we order
the $v_{ij}$ evaluations in ascending order, $v_{i(1)}\leq
v_{i(2)}\leq\ldots\leq v_{i(N_{\rm s})}$, and determine $v_i'$ through
a weighted mean

\begin{equation}
    v'_{i} = \sum_{j=1}^{N_{\rm s}}z_{(j)} v_{i(j)}, 
    \label{averaging}
\end{equation}
where $z_{(j)} \leq 1$ is the weight assigned to group $(j),\, \forall
j$. For the remainder of this discussion, we drop the $(\cdot)$
notation and assume that the $v_{ij}$ evaluations are listed according
to the ordering given above.  For the arithmetic mean, $z_{j}=1 /
N_{\rm s}$ and all groups are treated equally.  When the number of
groups $N_{\rm s}$ is odd, the median is obtained by setting
$z_{(N_{\rm s}+1)/2}=1$ and all other values $z_j =0$ in
Eq.\,\eqref{averaging}.  When $N_{\rm s}$ is even, the median is
obtained by setting $z_{N_{\rm s}/2}=z_{N_{\rm s}/2+1}=1/2$ and all
other values $z_{j} =0$.  Another possibility is to assume that the
most influential groups are, say, the most populous and assume that
$z_{j}$ is proportional to the number of constituents in group $j$.
In our work, we will utilize the Median and the two estimator methods
listed below, based on order statistics and on dropping or modifying a
portion of the evaluations $v_{ij}$.

\begin{itemize} 
    \item[(a)]{\textbf{Trimmed mean}: Under the assumption that $2 p <
      N_{\rm s}$, the input of $2 p$ out of the $N_{\rm s}$ groups is
      eliminated from the determination of $v_i'$. Of the discarded
      $v_{ij}$ evaluations, $p$ are the highest and the remaining $p$
      are the lowest ones. The intermediate $N_{\rm s} - 2p$
      evaluations $v_{ij}$ are kept and $v_i'$ is determined as their
      arithmetic mean. This scenario is called the $\alpha$-trimmed
      mean, where $\alpha = p/N_{\rm s}$. Mathematically,
    
    \begin{equation}
        z_{j}=
        \begin{cases}
            \displaystyle{\frac{1}{N_{\rm s}-2p}}  & \mbox{for }  p+1\leq j \leq N_{\rm s}-p,  \\
            \\
            \displaystyle{\quad \, 0} &\text{otherwise}. 
        \end{cases}
    \end{equation}}
\end{itemize}

\begin{itemize}     
    \item[(b)] \textbf{Winsorized mean}: This scenario is akin to the
      $\alpha$-trimmed mean. The only difference is that the $p$
      evaluations on the two tails are replaced with the nearest
      evaluations in the ordered range. In this case
    
    \begin{equation}
        z_{j}=
        \begin{cases}
            \quad \, 0  & \mbox{for } j \leq p \text{  or  } j \geq N-p+1, \\
            \\
             \displaystyle{\,\,   \frac{1}{N_{\rm s}}}  & \mbox{for } p+2\leq j \leq N-p-1,  \\
             \\
            \displaystyle{\frac{p+1}{N_{\rm s}}} &\mbox{for }  j=p+1 \text{  or  } j= N-p.  
        \end{cases}
    \end{equation}
    \end{itemize}
If only $N_{\rm s} = 3$ groups are present, the Trimmed and
Winsorized means are equivalent to each other and reduce to the
Median. This is the scenario in which groups agree to select the
intermediate evaluation among the three of them as their aggregated
value.

\subsubsection{Minimum variance:} 
In this case, the $v_{ij}$ values are weighted by project-specific
weights $z_{ij}$ that are larger for projects with smaller perception
errors $\sigma_{ij}$.  Thus, this aggregation process is biased in
favor of groups whose expertise is most closely aligned with project
type. We write
     
    \begin{equation}
    v'_i =   \sum_{j=1}^{N_{\rm s}}z_{ij} v_{ij};  \qquad 
    z_{ij} = \frac{r_{ij}}
  {\sum_{j=1}^{N_{\rm s}} r_{ij}}, 
    \label{linear_combination}
    \end{equation} 
where $r_{ij}$ is introduced to specify that the $z_{ij}$ weights are
normalized to unity.  A possible choice would be setting $r_{ij} =
\sigma_{ij}^{-\gamma}$, for $\gamma >0$. Since $v_{ij} = v_i +
\epsilon^{\rm p}_{ij}$ and the $\epsilon^{\rm p}_{ij}$ values are
derived from a Gaussian distribution of variance $\sigma_{ij}$, we can
calculate the total variance $\mbox{Var}(v'_{i})$ for the aggregated
$v_i'$.  If we assume that groups evaluate projects independently,
then
  
\begin{equation}
\label{linear_combination2}
\mbox{Var}(v'_{i}) = \sum_{j=1}^{N_{\rm s}} z^2_{ij} \sigma^2_{ij},
\end{equation}
which, for $\sigma_{ij} \neq 0$ for all $j$, and by imposing the
constraint $\sum_{j=1}^{N_{\rm s}} z_{ij} = 1$ through a Lagrange
multiplier, is minimized at $r_{ij} = \sigma_{ij}^{-2}$.  If
$\sigma_{ij} = 0$ for a given $j^*$ then we use $z_{\ij^*} = 1$ and
$z_{\ij^*} = 0$ for $j \neq j^*$, leading to $v_i' = v_{ij^*}$.
     
\subsubsection{Individual:}  Under perception noise, $\epsilon_{ij} = \epsilon^{\rm p}_{ij}$, 
the aggregated value of any project is determined by the evaluation
provided by the group with most centralized expertise. Effectively,
all groups agree to delegate evaluations to one group, regardless of
project type, so that
    
    \begin{equation}
    \label{ind}
        v'_i = v_{ik}, \qquad \mbox{where } \Big \rvert 
        \frac{t_{\rm min}+t_{\rm max}}{2} - e_{k} \Big \rvert \leq \Big \rvert \frac{t_{\rm min}+t_{\rm max}}{2} - e_j \Big \rvert,\forall j.
    \end{equation}
The group with expertise $e_k$ selected via Eq.\,\eqref{ind} may not
be particularly qualified to evaluate any specific project $i$, but
its generalist knowledge may be sufficient to guarantee an overall
evaluation of the long-term benefits of the projects at hand.  In this
case, the final decisions reflect the sole priorities of group $k$.
The earlier assumptions made in this paper on project type and group
expertise ranges and listed in Table\,\ref{tab:model_quantities} allow
us to identify $k$ with the group that carries the average expertise,
so that $e_k = e_{\rm M} = t_{\rm M}$. This is not the case in more
general settings.

\subsubsection{Delegation:} 
This case is similar to the Individual method, with the difference
that the aggregated value of each project is determined by the group
whose expertise is most closely aligned with the corresponding project
type.  Effectively, given a project $i$, all groups agree to utilize
the evaluation made by the group $k$ that has the best understanding
of the project.  The underlying assumption is that of a relatively
cohesive society, where there are no specific benefits or detriments
that a project can confer to one group over another, and where there
is a consensus that one group may have better skills than others in
evaluating a given project. We write
    
\begin{equation}
\label{delegate}
 v'_i = v_{ik}, \qquad \mbox{where } |t_i - e_{k}|\leq |t_i - e_j|, \,
 \forall j.
\end{equation}
Note that since $v_{ij}$ is selected from a Gaussian distribution, a
group with better expertise may still assign to a given project a
higher or lower value compared to a group with lesser expertise.  This
method was proposed by \cite{csaszar2013organizational} for
single-project evaluation and later expanded by
\cite{boettcher2024selection} to a portfolio context.

\subsection{Indirect aggregation methods}
\label{indirect}

In the above section, we derived the aggregated value $v_i'$ of
project $i$ through direct manipulation of the group evaluations
$v_{ij}$. Here, we introduce a new aggregation attribute of project
$i$, its ``score" $q'_i$.  This quantity is derived from the $v_{ij}$
evaluations but its magnitude may be very different both from the
$v_{ij}$ evaluations and the aggregated value $v_i'$.  The objective
of introducing a score is to provide a concise and easily
interpretable measure of a project's desirability.  In this paper, we
determine the $q'_i$ scores as follows. Each $j$ group is asked to
determine the quality $q_{ij}$ of each project, where $q_{ij} \equiv
v_{ij} / w_i$ is the value to cost ratio of project $i$.  For fixed
$j$, we can then order the $q_{ij}$ qualities so that the resulting
$i$-orderings, say, list the highest to lowest quality project for
group $j$.  If all costs are the same, then $w_i = w$ and the $v_{ij}$
and $q_{ij}$ $i$-orderings are the same for each $j$; in the more
general case of heterogeneously distributed costs, the $i$-orderings
as determined by group $j$ may differ depending on whether one chooses
values or quality.

Once the $q_{ij}$ qualities are determined, all $i$ projects are assigned an 
overall quality $q'_i$, using the $q_{ij}$ inputs from all groups 
and one of the different aggregation methods discussed below. 
The $q'_i$ quality is the score of the project.  

The aggregation methods we use typically result in scores that are
integer numbers, or that are distributed within easily interpretable
intervals, yielding an immediate understanding of whether project $i$
is desirable or not.  Finally, the collective knapsack is filled by
optimizing the following problem

\begin{equation}
    \max \sum_{i=1}^{N_{\rm p}} q'_i w_ix_i, 
    \label{max2}
\end{equation}
subject to the constraint in Eq.\,\eqref{constraint}. This formulation of the knapsack problem is equivalent
to the original one in Eq.\,\eqref{max2} upon defining $q_i' w_i = v_i'$. 

Some indirect aggregation methods involve rescaling the original
$q_{ij}$ values so that the range and mean of the aggregated $q_i'$
scores are fixed or mapped onto preset intervals. One well known example
is the $z$-score where all values of a given distribution are mapped
onto the standard Gaussian. Once the target distribution or interval is determined,
the $q_{ij}$ values are mapped onto them through a proper 
transformation $T_{\rm s}$. 

 \subsubsection{Borda count:}
 
 Here, each group ranks projects by their quality $q_{ij} = v_{ij}/ w_i$
 so that  $q_{(1),j}\geq q_{(2),j}\geq\ldots\geq q_{(N_{\rm p}),j}$. According to this ordering, group $j$ determines that 
 project $(1)$ has the best quality. 
 An integer number $0 \leq k_{i,j} \leq N_{\rm p}-1$ is now assigned to each project 
according to the above ranking so that the project with highest quality is labeled by the highest $k_{i,j}$ integer.
The $k_{i,j}$ integers are also referred to as points. In the example above, project $(1)$, with the highest quality $q_{(1),j}$ is assigned
$k_{(1),j} = N_{\rm p}-1$ points, whereas project $(N_{\rm p})$, with the lowest quality $q_{(N_{\rm p}),j}$, is assigned 
$k_{(N_{\rm p}),j }= 0$ points.
The points each project receives from all groups
are then added,  yielding the Borda count $s_i$. The project with the highest $s_i$ is commonly known as the
``winner".  Mathematically, the Borda count $s_i$ for project $i$ is 

\begin{table}[t]
\centering
\begin{tabular}{|c|c|c|c|c|c|c|}
\hline
& \textbf{Group 1} & \textbf{Group 2} & \textbf{Group 3} & \textbf{Group 4} & \textbf{Group 5} & \textbf{Total Borda count} \\
\hline
\textbf{Project A } & 2 & 1 & 1 & 2 & 0 & $s_{\rm A} = 6$ \\
\textbf{Project B } & 1 & 2 & 0 & 0 & 2 & $s_{\rm B} = 5$ \\
\textbf{Project C } & 0 & 0 & 2 & 1 & 1 & $s_{\rm C} = 4$\\
\hline
\end{tabular}
\vspace{0.3cm}
\captionsetup{justification=justified}
\vspace{0.2mm}
\caption{
Example of the Borda count for $N_{\rm p} = 3$ projects labeled A,B,C and five groups, $N_{\rm s} = 5$. 
Each entry represents the $k_i^j$ points assigned to project $i$ by group $j$.  
For example, group 1 determines project A  to have the best quality, project B to have intermediate quality,
and project C to have the least quality, resulting in $k_{\rm A}^1 =2$  points, $k_{\rm B}^1 =1$  point, $k_{\rm C}^1 =0$  points. 
The 
maximal Borda count is $s_{\rm max} = N_{\rm s} (N_{\rm p} -1) = 10$; the minimum is 
$s_{\rm min} = 0$. The rankings
shown in this voting system result in project A (corresponding
to $i=1$) being the winner with total Borda count 
$s_{\rm A} = 2 +  1 + 1 +1 + 0 = 6$. The runner up is  project B ($i=2$)
with total Borda count 
$s_{\rm B} = 1 +  2 + 0 +0 + 2 = 5$. The lowest priority project is 
C ($i=3$) with total Borda count of 
$s_{\rm C} = 0 +  0 + 2 + 1+ 1 = 5$. We use the scores $q'_i = s_i$ to fill the knapsack
described in Eqs.\,\eqref{max2} and \eqref{constraint}.
}
\label{bordatable}
\end{table}
  
  \begin{equation}
      s_i=\sum_{j=1}^{N_{\rm s}} k_{i,j}; \qquad q'_i = s_i, \quad   q'_i \in \{0,  \dots,  N_{\rm s} (N_{\rm p}  -1) \}. 
  \end{equation}
If all groups rank project $i$ as their most preferred, then its Borda count is also the maximum attainable,  
$s_i = s_{\rm max} = N_{\rm s}(N_{\rm p} - 1)$; whereas if all groups rank project $i$ the lowest, then its Borda count
is also the lowest, $s_i = s_{\rm min} =0$. 
The Borda count is a consensus-based voting method and not a majoritarian one
since the overall winner may be broadly acceptable to a large number of groups, without being the 
one ranked first by the majority of them.  
Once calculated, we identify the Borda count $s_i$ for project $i$ with the score
$q'_i$ in Eq.\,\eqref{max2} and proceed to solve the knapsack problem
given by Eqs.\,\eqref{max2} and \eqref{constraint} with $q'_i = s_i$. 
Note that neither the cost of each project $w_i$ nor the total budget $W$
are affected by the Borda count. 

We illustrate the Borda count in Table\,\ref{bordatable} 
with $N_{\rm p} = 3$ projects, and $N_{\rm s} = 5$ groups. 
Within this setup, scores can vary within the interval
$0 \leq s_i \leq 10$. 
We label the three
projects as
$A$ (corresponding to $i=1$),  
$B$ (corresponding to $i=2$), 
$C$ (corresponding to $i=3$) and assume the rankings are as follows: 
for group $j=1$,  $q_{A,1} > q_{B,1} > q_{C,1}$; 
for group $j=2$,  $q_{B,2} > q_{A,2} > q_{C,2}$;
for group $j=3$,  $q_{C,3} > q_{A,3} > q_{B,3}$;
for group $j=4$,  $q_{A,4} > q_{C,4} > q_{B,4}$;
for group $j=5$,  $q_{B,5} > q_{C,5} > q_{A,5}$.
The total Borda count (and the score) for each project 
is given by $q'_{\rm A} = s_{\rm A} = 6, q'_{\rm B} =  s_{\rm B} = 5, q'_{\rm C} =  s_{\rm C} = 4$
as shown in Table \,\ref{bordatable}, 
and the winner is project $A$.  The $q'_i$ scores can be now used to solve the knapsack problem 
in Eqs.\,\eqref{max2} and \eqref{constraint}.

 \subsubsection{Yes-or-no voting:} Under this voting scheme, each group expresses a  ``yes'' or ``no'' preference on each project, depending
on how its evaluation $v_{ij}$ compares to an internal cutoff. In principle, this cutoff may be different among groups,
so that although two different groups may evaluate the same project in the same way, one 
may result in a yes vote,  the other in a no vote.
We then tally all positive votes for each project and utilize this quantity as the score $q'_i$.  
If all the $N_{\rm s}$ groups vote yes on a project, it will receive the maximum score, $q'_i = q'_{\rm max}= N_{\rm s}$;
if they all vote no, the project be assigned the minimum score $q'_i = q'_{\rm min} = 0$.  
We impose the internal cutoff to be zero, uniformly for all groups,  thus
project $i$ is assigned a ``no" vote if $v_{ij} \leq 0$ and a ``yes" vote if 
$v_{ij} > 0$ for all $j$.  We write

\begin{equation}
q'_i = \sum_{j=1}^{N_{\rm s}} \mathbf{1}_{Y}(j); \quad  q'_i \in \{0,  \dots,  N_{\rm s} \}. 
\end{equation}
Here, $\mathbf{1}_{Y}(j) = 1$ if group $j$ expresses a yes vote, and $\mathbf{1}_{Y}(j) = 0$  if group $j$ expresses a no vote.
We can now proceed with the knapsack problem given in Eqs.\,\eqref{max2} and \eqref{constraint}.

 \subsubsection{Min-max scaling:}
This is a scaling method whereby the original $q_{ij}$ qualities
are mapped onto new variables that are distributed in the $[0,1]$ interval. 
To do this, groups first identify the two projects they 
evaluated as having the largest and smallest $q_{ij}$.
We denote them as $q_{{\rm max}, j} \equiv {\rm{max}}_i  \, \{ q_{ij} \}$ and 
$q_{{\rm min}, j}  \equiv {\rm{min}}_i  \,  \{ q_{ij} \}$ and use them as endpoints
to create the  $[0,1]$ interval. The linear transformation $T_{\rm s}$
that maps all other $q_{ij}$ to the $[0,1]$ interval  is 

\begin{equation}
    T_{\rm s}(q_{ij})=\frac{q_{ij}-q_{{\rm min}, j}}{q_{{\rm max}, j} - q_{{\rm min}, j}}, 
    \end{equation}
    so that $T_{\rm s}(q_{{\rm max}, j}) = 1$, and $T_{\rm s}(q_{{\rm min}, j}) = 0$. The score for project $i$ 
    is now determined by adding the values $T_{\rm s}(q_{ij})$ across all groups $j$
    
\begin{equation}
    q'_i = \sum_{j=1}^{N_{\rm s}} T_{\rm s}(q_{ij});  \quad q'_i \in \{0,  \dots,  N_{\rm s} \}. 
    \end{equation}

 \subsubsection{z-score scaling:}
One common method to rescale the $q_{ij}$ qualities is to derive their 
$z$-score.  If $\mu_j$ and $s_j$ denote, respectively, 
the mean and standard deviation obtained from averaging the $q_{ij}$ qualities over all 
$N_{\rm p}$ projects, the $z$-score $z(q_{ij})$ is 

\begin{equation}
\label{zscore}
z(q_{ij})=\frac{q_{ij}-\mu_j}{s_j};  \qquad \mu_j = \frac 1 {N_{\rm p}} \sum_{i=1}^{N_{\rm p}} q_{ij} ; 
\qquad  s^2_j = \frac 1 {N_{\rm p}} \sum_{i=1}^{N_{\rm p}} (q_{ij} - \mu_j)^2. 
\end{equation}
For all $N_{\rm s}$ groups, the mean determined by the $z$-score rescaling is zero, 
the standard deviation is unity, and the range of the rescaled qualities is the entire real axis, allowing for meaningful comparisons across
groups.  The transformation $T_{\rm z}(q_{ij}) = z(q_{ij})$  yields the $q_i'$ score as

\begin{equation}
\label{zscoreavg}
    q'_i = \sum_{j=1}^{N_{\rm s}} T_{\rm z}(q_{ij}) =  \sum_{j=1}^{N_{\rm s}} z(q_{ij}) ;  \quad  q'_i \in (-\infty, \infty).
    \end{equation}    
However, the transformation in Eq.\,\eqref{zscore} yields a negative $z$-score for
approximately half the projects,  those whose perceived quality $q_{ij}$ is less than the mean $\mu_j$. 
This could lead to a large number projects with negative $q'_i$ scores,  unless elaborate transformations are included 
in the definition of $q'_i$. 
Projects with negative $q_i'$ are not included in the knapsack, as they decrease the value
of the knapsack content which we seek to maximize as
per Eq.\,\eqref{max2}. Although the definition of $v_{ij}$ (and hence of $q_{ij}$)
as the superposition of a positive $v_i$ and a random quantity $\epsilon^{\rm p}_{ij}$ selected from a Gaussian, 
does allow for some negative $v_{ij}$ values, the $z$-score metric in Eq.\,\eqref{zscore}
is designed to systematically yield a negative result on about half the entries. 
We thus predict it will not be an efficient aggregation method.

 \subsubsection{Standard-deviation scaling:}
 This is a scaling method designed to improve on the $z$-score scaling. 
Here, the quality of each project is weighted by the standard deviation $s_j$ 
as defined in Eq.\,\eqref{zscore} but not relative
to the average $\mu_j$ so that 

\begin{equation}
\label{zscore2}
    T_{\rm z}(q_{ij})=\frac{q_{ij}}{s_j}. 
\end{equation}
This transformation results in all groups having the same (unitary) standard deviation and their rescaled qualities being
distributed along the real axis. However the corresponding mean is not the same across groups. 
However, Eq.\,\eqref{zscore2} also leads to a knapsack with higher value than the $z$-score aggregation
method as there are less projects discarded, so we include it as a possible scaling method.
Similarly to other scaling methods,  we define the $q'_i$ score via

\begin{equation}
\label{zscoreavg2}
    q'_i = \sum_{j=1}^{N_{\rm s}} T_{\rm z}(q_{ij});  \quad  q'_i \in (-\infty, \infty).
    \end{equation}    
Although the $q'_i$ qualities determined via Eq.\,\eqref{zscore2} may be negative,
the likelihood of this occurring is much lower than under $z$-score
since we do not map the $q_{ij}$ qualities onto negative rescaled values. 

\subsection{Applicability to societal decision-making}

The twelve aggregation methods discussed above can be catalogued by how representative and/or transparent they are.
The Trimmed mean,  Individual and Delegation methods for example,  utilize only the $v_{ij}$ estimations of select groups in determining the
aggregated value $v_i$ of project $i$, with other groups serving as a reference. Within this context, the 
Individual and Delegation approaches are the most representative methods, since aggregation results in delegating
to the judgement of a single reference  group.  Other methods, such as the Arithmetic mean, 
allow for all groups to partake equally in the aggregation outcome. Selecting one method over the other involves a variety of 
considerations. For example, some groups may recognize their lack of expertise and voluntarily defer to the decisions
of others. Similarly, during a time of emergency there might not be enough time or resources to engage in extensive 
discourse and only those with a specialized understanding of the subject may be queried. In other circumstances, there may be
complex technical or political factors to consider, including safety or societal stability that may require prioritizing
the input of specific groups to mitigate potential negative consequences. 
However, as a general rule, it is important to prioritize inclusivity and to ensure that all evaluations are properly taken into account. 

Finally, some methods are associated with a high degree of transparency and sharing of information 
compared to others. For example, the Minimum variance and Delegation methods require 
that groups disclose their level of expertise so that their evaluations
may be weighted properly.  Other methods are more private and less information is necessary
to determine $v_i'$. 
We provide a schematic in Fig.\,\ref{fig:social_classification_demo}.

\begin{figure}[t!]
    \centering
\begin{tikzpicture}

\tikzstyle{every node}=[font=\small]

\draw[thick] (0,0) rectangle (8,8);
\draw[thick] (4,0) -- (4,8);
\draw[thick] (0,4) -- (8,4);

\node at (2,8.5) {\bf Privacy};
\node at (6,8.5) {\bf Transparency};
\node[rotate=90] at (-0.5,6) {\bf Representative};
\node[rotate=90] at (-0.5,2) {\bf Direct democracy};

\node[align=center] at (2,6) {
  Median\\
  Trimmed mean\\
  Winsorized mean
};

\node[align=center] at (6,6) {
  Delegation\\ 
  Individual
};

\node[align=center] at (2,2) {
  Arithmetic mean\\
  Borda count\\
  Yes-or-no voting\\
  Min-max scaling\\
  z-score\\
  Standard-deviation scaling
};

\node[align=center] at (6,2) {
  Minimum variance
};

\end{tikzpicture}
    \captionsetup{justification=justified}
    \caption{The proposed twelve aggregation methods catalogued by level of transparency and democracy type. 
     The horizontal axis parts representative methods (whereby decisions depend on the input of a select number of chosen groups) 
    from direct methods (whereby decisions depend on the input of all groups). The vertical axis parts 
     methods marked by high levels of transparency (where information from single groups is shared with the centralized agency 
     that is responsible for the final decision) from private methods (where less information is shared and only aggregate results
     are needed for final decisions to be made).  The aggregated values $v_i'$ for $i \in \{1, \dots, N_{\rm p} \}$ obtained from any of these methods 
    are then used to fill the collective knapsack of budget $W$. } 
    \label{fig:social_classification_demo}
\end{figure}

\section{An illustrative example with $N_{\rm p} =2$ projects}
\label{sec:illustrative}
As a first example, we consider a simple selection problem under perception noise, $\epsilon_{ij} = \epsilon^{\rm p}_{ij}$, 
when there are only two projects to choose from, of cost $w_1=w_2=1$
and value $v_1=a > 0$ and $v_2=b> 0$, with $a<b$ without loss of generality. We set the 
knapsack budget $W=1$ so that only one of the two projects can be selected. 
Finally, we let project types $t_1$ and $t_2$ be taken from
the uniform distribution $\mathcal{U}(t_{\rm min},t_{\rm max})$, 
with $t_{\rm min} < t_{\rm max}$.

We determine the aggregate values $v_1'$ and $v_2'$ for the $N_{\rm p} = 2$ projects using the
following aggregation rules: (i) Arithmetic mean, (ii) Individual,  and (iii) Delegation. 
We also discuss (iv) the Median when $N_{\rm s} = 3$. 
The latter case is the same as the Trimmed and Winsorized means for $N_{\rm s} = 3$, and represents the case where 
groups agree to select the intermediate evaluation among them as their aggregated value.  
We select these aggregation methods as they are amenable to analytical treatment for  $N_{\rm p} = 2$. 
The remaining methods, or 
scenarios with  $N_{\rm p} > 2$ projects, will be examined in the next section 
using numerical methods. 

If we assume that projects are evaluated independently of each other, by construction, 
methods (i)--(iv) lead to aggregate values $v_1'$ and $v_2'$ 
that are independent random variables normally distributed with mean $v_1=a$ and $v_2=b$, respectively. 
The corresponding standard deviations $g_{\rm agg}(t_i)$ depend on the aggregation method, 
the project type $t_i$ for $i \in \{1,2 \}$, the number of groups $N_{\rm s}$, 
and the group expertise distribution parameters $e_{\rm M}, \beta.$
We write

\begin{equation}
\label{random1}
    v_1'\sim \mathcal{N}(a, g_{\rm agg}(t_1)) \quad \mathrm{and} \quad
    v_2'\sim \mathcal{N}(b,g_{\rm agg}(t_2)),
\end{equation}
where $g_{\rm agg} (t_i) =g_{\rm avg}(t_i)$,  $g_{\rm agg} (t_i) = g_{\rm ind} (t_i)$,  $g_{\rm agg} (t_i) = g_{\rm del}(t_i)$,
or $g_{\rm agg} (t_i) = g_{\rm med}(t_i)$ for each of the
(i)--(iv) aggregation methods described above. We evaluate them below.  
In all cases, we use the standard deviation associated to perception noise $\sigma_{ij} = |t_i - e_j|$
and  refer to $\sigma_{ij}$ as the perception error. 
Once the values of $g_{\rm agg}(t_i)$ are determined, we calculate
$V(t_1, t_2)$, the expected value of the filled knapsack with the constraint that
only one project can be selected as

\begin{equation}
\label{vknap}
V(t_1, t_2) = \mathbb{E} \left[ \sum_{i =1}^ {N_{\rm p}=2  }x_i v_i \right]. 
\end{equation}
The expectation in Eq.\,\eqref{vknap} is determined over many realizations 
of the $N_{\rm s}$ groups evaluating the same two projects of type $t_1, t_2$. 
Although we only highlight the dependence
on $t_1, t_2$ in Eq.\,\eqref{vknap},  $V(t_1, t_2)$ also depends on the values $a,b$, on the number
of stakeholder groups $N_{\rm s}$, and on the knowledge breadth $\beta$. 
Due to the stochasticity of the evaluation process, $x_i \in \{0,1 \}$ is a random variable;
due to the budget constraints the only two possible combinations are
$x_1 = 1, x_2 =0$ or $x_1 = 0, x_2 =1$. 
Given the aggregated values $v'_1$ and $v'_2$, and recalling that the filled knapsack can have value $v_1 = a$
with  $x_1 = 1, x_2 =0$ or $v_2 = b$ with $x_1 = 0, x_2 =1$
depending on the relationship between $v'_1$ and $v'_2$, we write
\begin{align}
\begin{split}
\label{valuetwo}
    V(t_1,t_2)&=a \Pr(v_1'\geq v_2')+ b \Pr(v_1'< v_2')  \\
    &=a(1-\Pr(v_1'< v_2'))+b\Pr(v_1'< v_2')\\
    &=a+(b-a)\Pr(v_1'- v_2'< 0), 
\end{split}
\end{align}
where $\Pr(\cdot)$ denotes the probability, and the last equality follows from well know identities in probability theory,
{\textit i.e.} $\Pr (x < y) =\Pr (x -y < 0) $.  Using Eq.\,\eqref{random1} it is straightforward to show that 
the quantity $v_1'-v_2'$ satisfies
\begin{equation}
	v_1'-v_2' \sim \mathcal{N} \left( a-b,\sqrt{g^2_{\rm agg}(t_1)+g^2_{\rm agg}(t_2)} \right).
\end{equation}
We thus obtain
\begin{equation}
\Pr(v_1'-v_2'<0)=\Phi\left(\frac{b-a}{\sqrt{g^2_{\rm agg}(t_1)+g^2_{\rm agg}(t_2)}} \right), 
\end{equation}
where $\Phi(z)$ is the cumulative distribution function of the standard normal distribution $\phi(t)$ defined as

\begin{equation}
\label{stdgauss}
\Phi(z) \equiv \int_{-\infty}^{z} \phi(t) dt =  
\frac 1 {\sqrt{2 \pi}} \int_{-\infty}^z e^{-\frac{t^2}{2}} dt. 
\end{equation}
Finally, we write
\begin{equation}V(t_1,t_2)=a+(b-a)\Phi\left(\frac{b-a}{\sqrt{g^2_{\rm agg}(t_1)+g^2_{\rm agg}(t_2)}}\right).
\label{pdf_of_performance}
\end{equation}
Since $0 \leq \Phi(z) \leq 1$, and $\Phi(z)$ is an increasing function of $z$, and since $g_{\rm agg}(t_i)$ is independent
of $a,b$,  it follows that $a \leq V(t_1, t_2) \leq b$
and that $V(t_1, t_2)$ increases as a function of $b-a > 0$ for all aggregation methods. We now determine
$g_{\rm agg}(t_i)$ for the aggregation methods (i)--(iv) and discuss the emerging trends in
$V(t_1, t_2)$.

\begin{itemize}
\item[(i)]{\textbf{Arithmetic mean:}
Using well known properties of the mean of random variables that are normally distributed about a common
center, we write $g_{\rm avg}(t_i)$ for the Arithmetic mean as
\begin{equation}
\label{gavg}
    g_{\rm avg}(t_i) = \frac 1 {N_s} 
    \left(\sum_{j=1}^{N_{\rm s}} \sigma^2_{ij}\right)^{1/2} \qquad {\mbox {where }}  \sigma_{ij} = |t_i - e_j|. 
\end{equation}}

It can easily be shown using Eq.\,\eqref{expert} that $V(t_1, t_2)$ is 
a decreasing function of $\beta$ for fixed $N_{\rm s}$. Increasing
the heterogeneity $\beta$ of the group expertise levels leads to
less accurate assessments of projects. This, in turn, lessens $V(t_1, t_2)$ since all inputs are used to determine 
the final selection. It can be similarly shown that, for fixed
$\beta$, $V(t_1, t_2)$ is an increasing function of $N_{\rm s}$. 

\end{itemize}
\begin{itemize}
\item[(ii)]{\textbf{Individual:}
For this protocol,  since $v'_i$ is determined by delegating
to a specific group $k=k^*$ as per Eq. \eqref{delegate},  the associated standard deviation is given by 
\begin{equation}
\label{gind}
    g_{\rm ind}(t_i) = \sigma_{ik^*} =  |t_i - e_{k^*}|, 
     \qquad \mbox{where \,}  
      \rvert  t_{\rm M} - e_{k^*}
      \rvert  \leq \rvert  t_{\rm M}- e_j \rvert,  \, \forall j
\end{equation}
and where $t_{\rm M} = (t_{\rm min} + t_{\rm max})/2$ is the center of the project-type interval.
In this case, $k^*$ is independent of $i$ and $g_{\rm ind}(t_i)$ viewed as a function of
$t_i$ is the absolute value of $t_i$ shifted with respect to $e_{k^*}$. 
It is straightforward to conclude that 
$V(t_1, t_2)$ decreases if the expertise of the representative group $k^*$  is increasingly divergent
from the project types at hand.  Since we assume the center of the expertise and project-type intervals coincide, 
{\textit {i.e.}} $e_{\rm M} = t_{\rm M}$, Eq.\,\eqref{gind} implies that the group $k^*$ to delegate decisions to
is the one with the central expertise, $e_{k^*} = e_{\rm M}$. 
This is true for all $N_{\rm p}, \beta$ and for odd $N_{\rm s}$. 
}
\end{itemize}

\begin{itemize}
\item[(iii)]{\textbf{Delegation:}
The standard deviation under the Delegation aggregation protocol is given by the piece-wise linear function
\begin{equation}
\label{gdel}
g_{\rm del}(t_i)= \sigma_{ik(i)} =  |t_i-e_k (t_i)|,  \qquad {\mbox {where }} |t_i - e_k(t_i)|\leq |t_i - e_j|, \, \forall j.
\end{equation}
Here, $g_{\rm del}(t_i)$ is the absolute value of $t_i$ shifted by an amount, $e_k(t_i)$, that depends on $t_i$ itself.
Similarly to the Individual aggregation method,  $V(t_1, t_2)$ decreases as
the expertise of the group whose judgement all others have delegated to (to evaluate project $i$)
diverges from the project type $t_i$. How $N_{\rm s}, \beta$ affect $V(t_1, t_2)$
depends on how the distribution of the expertise levels compares to the $t_i$ project types. 
}
\end{itemize}

\begin{itemize}
\item[(iv)]{\textbf{Median:}
Finally, upon setting $N_{\rm s} = 3$ and ordering the respective evaluations in ascending order, 
the Median aggregation method yields

\begin{equation}
g_{\rm med}(t_i)= \sigma_{i(2)} = |t_i - e_{(2)}|  \qquad {\mbox {where }} v_{i(1)} \leq v_{i(2)} \leq v_{i (3)},  \, \forall j, 
\end{equation}
which for $N_{\rm s} = 3$ is the same as the Trimmed and Winsorized mean. 
We can now calculate

\begin{equation}
\label{asymm1}
\Pr(v_1' - v_2' < 0) = 
\sum_{\substack{k \neq \ell \neq  m}}^{N_{\rm s}}
\sum_{r\neq s\neq t}^{N_{\rm s}}
\Pr(v_{1 \ell}-v_{2s}<0 \,  | (v_{1k} \leq v_{1\ell} \leq v_{1 m} ) 
{\cap} (v_{2r} \leq v_{2 s} \leq v_{2 t} )), 
\end{equation}
where the quantity $\Pr(v_{1 \ell}-v_{2s}<0 \,  | (v_{1k} \leq v_{1\ell} \leq v_{1 m} ) 
{\cap} (v_{2r} \leq v_{2 s} \leq v_{2 t} ))$ 
 is the probability that  $v_{1\ell} < v_{2s}$ conditioned on 
the $v_{1 \ell}, v_{2 s}$ evaluations of
projects $i=1$ and $i=2$ made by groups $\ell, s$, 
respectively, lie between the corresponding 
evaluations made by the other groups.  
Under these assumptions, by construction 
$v'_1 = v_{1\ell}$ and $v'_2 = v_{2s}$. 
Given a set of three independent random variables $X_1,Y_1,Z_1$
distributed according to $f_{X_1}(x_1)$, $f_{Y_1}(y_1)$,  $f_{Z_1}(z_1)$, 
and another set $X_2,Y_2, Z_2$
distributed according to $f_{X_2}(x_2)$, $f_{Y_2}(y_2)$,  $f_{Z_2}(z_2)$,
respectively, we can write

\begin{equation}
\begin{aligned}
\label{concatenate}
\Pr(Y_1 < Y_2 | & (X_1 \leq Y_1 \leq Z_1) \cap  (X_2 \leq Y_2 \leq Z_2))  = \\
& \int \int_{y_2 > y_1}   f_{Y_1} (y_1) f_{Y_2} (y_2) \, 
F_{X_1}(y_1) (1 - F_{Z_1}(y_1))  F_{X_2}(y_2)   (1-F_{Z_2}(y_2))  \, dy_1  dy_2, 
 \end{aligned}
\end{equation}
where the first integral is over the entire domain of $f_{Y_1}(y_1)$
and the second is restricted to values  $y _2 >  y_1$ in the domain
of $f_{Y_2}(y_2)$.   In Eq.\,\eqref{concatenate}
$F_{X_i} (y_i)$ is the cumulative distribution 
function associated to $f_{X_i} (x_i)$ defined as
\begin{equation}
F_{X_i} (y_i) = \Pr(X_i \leq y_i) = \int_{x_i < y_i} f_{X_i} (x_i) dx_i, 
\end{equation}
and similarly for $F_{Z_i} (y_i)$. 
Since all evaluations $v_{ij}$ are derived from
 Gaussian distributions, we can write 
 Eq.\,\eqref{asymm1} as

\begin{equation}
\begin{aligned}
\label{asymm3}
\Pr(v_1' - v_2' < 0)  = 
 \int \int_{y_2 > y_1}   \sum_{\substack{k \neq \ell \neq  m}}^{N_{\rm s}}
\sum_{r\neq s\neq t}^{N_{\rm s}}
 &   \phi \left(\frac{y_1- a}{\sigma_{1\ell}}  \right) 
\phi \left(\frac{y_2- b}{\sigma_{2s}}  \right) 
\Phi \left(\frac{y_1 -a}{\sigma_{1k}}\right)
 \Phi \left(\frac{y_2 -b}{\sigma_{2r}}\right) \times \\
& \left[1- \Phi \left(\frac{y_1 -a}{\sigma_{1m}}\right) \right]
\left[1- \Phi \left(\frac{y_2 -b}{\sigma_{2t}}\right) \right]
dy_1  dy_2, 
\end{aligned}
\end{equation}
\noindent
where $\Phi(\cdot)$ and $\phi(\cdot)$ are defined in Eq.\,\eqref{stdgauss}. 
}
\end{itemize}
Finally, we introduce the performance $E_2(\beta, N_{\rm s})$, defined as the average of the 
expected  value $V(t_1, t_2)$ taken over all possible project types $t_1, t_2$.
That is,

\begin{equation}
E_2(\beta, N_{\rm s}) = \frac{1}{(t_{\rm max} - t_{\rm min})^2} \int_{t_{\rm min}}^{t_{\rm max}}\int_{t_{\rm min}}^{t_{\rm max}} V(t_1,t_2)\,\mathrm{d}t_1\mathrm{d}t_2.  
\label{performance_expectation}
\end{equation}
This quantity is often used in the literature on organizational decision-making
as a metric that is independent of project type, so that results do not depend
on the specific choices made for $t_i$.  It is called performance in reference
to the performance of an investment portfolio of a firm, 
whereby different managers are called to decide which investments to fund. 
If the aggregation protocol is such that 
 $V(t_1, t_2)$ is independent of $t_1, t_2$ and reaches its maximal value 
 $V_{\rm max} (2) = b$, independently of $t_1, t_2$, then 
 $E_2(\beta, N_{\rm s}) = E_{ \rm max} (2)= V_{\rm max} (2)= b$. 
Of course, $E_2(\beta, N_{\rm s})  \leq V_{ \rm max}(2)= b$ for all aggregation methods. 

\begin{figure}
	\centering
	\includegraphics[width=\linewidth]{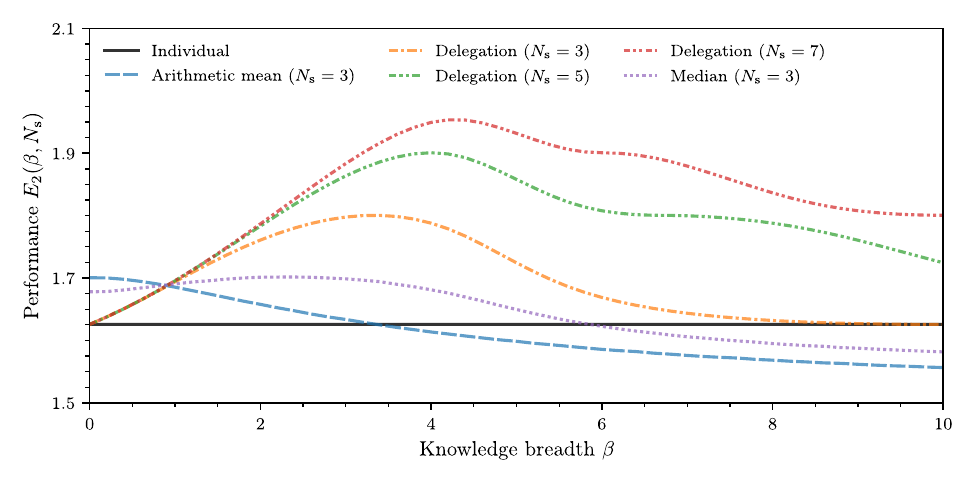}
        \captionsetup{justification=justified}
\caption{Performance $E_2(\beta, N_{\rm s})$ as defined in Eq.\,\eqref{performance_expectation} for  $N_{\rm p} =2$ projects of value $v_1 = a =1, v_2 = b=2$ and of 
cost $w_1=w_2=1$, respectively.  Group expertises are within $[e_{\rm M} - \beta, e_{\rm M} + \beta]$ where $e_{\rm M} = 5$ and $\beta$
can vary. Project types are uniformly distributed in 
$[t_{\rm min} , t_{\rm max}]$ where $t_{\rm min}=0, t_{\rm max} =10$ and the collective knapsack budget is set at
$W=1$, implying that only one project can be selected. The performance $E_2(\beta, N_{\rm s})$ is the average over all possible types $t_1, t_2 \in [t_{\rm min}, t_{\rm max}]$ 
of the expected value of the collective knapsack $V(t_1, t_2)$. The latter is obtained as the expectation of the evaluation process from the input of
$N_{\rm s}$ groups for fixed project types $t_1, t_2$. For the chosen parameters, 
$ E_{\rm max} (2) = V_{\rm max} (2) = b =2$. 
The aggregation methods surveyed are Individual, Arithmetic mean,  Delegation and the Median.  The curve corresponding
to the Arithmetic mean is a decreasing function
of $\beta$ and an increasing function of $N_{\rm s}$. The curves corresponding to Delegation 
are increasing in $N_{\rm s}$ for fixed $\beta$. For fixed $N_{\rm s}$, they 
display maxima at $\beta = \beta_{\rm opt}(N_{\rm s})$ estimated as  $\beta_{\rm opt} (3) = 10/3 = 3.33$, 
$\beta_{\rm opt} (5) = 4$, 
$\beta_{\rm opt} (7) = 30/7 =4.29$ as discussed in the text and in good agreement with numerical findings.
The curve corresponding to the Median for $N_{\rm s}$ is also non-monotonic. 
However, the estimated 
performance $E_2(\beta, N_{\rm s})$ is lower than under Delegation.}
 \label{fig: two_project}
\end{figure}

In Fig.\,\ref{fig: two_project}, we plot the performance $E_2(\beta, N_{\rm s})$ for 
the Individual, Arithmetic mean, Delegation and Median aggregation protocols
as a function of the knowledge breadth $\beta$ for various values of $N_{\rm s}$.
The parameters selected for the two projects are $a=1, b=2$ and $w_1 =w_2 =1$. 
The available budget is set at $W=1$ so that only one of them can be selected. 
We also set $t_{\rm min} =0$, $t_{\rm max} =10$, $e_{\rm M} = 5$ and vary $\beta$
and $N_{\rm s}$. 

The Individual method yields a uniform value for $E_2(\beta, N_{\rm s})$, 
independent of $N_{\rm s}$ and $\beta$. This is expected since
evaluations are delegated to group $k^*$ as per Eq.\,\eqref{gind}, 
and since in our specific setting ({\textit{i.e.}} $t_{\rm M} = e_{\rm M}$ and $N_{\rm s}$ odd) 
$e_{k^*} = e_{\rm M} = 5$ regardless of the number of stakeholder groups and their expertise spread.  
Using the above parameters and Eqs.\,\eqref{pdf_of_performance} and \eqref{gind}, 
a direct numerical integration of Eq.\,\eqref{performance_expectation} yields $E_2(\beta, N_{\rm s}) = 1.626$. 

The Arithmetic mean method yields $E_2(\beta, N_{\rm s})$ curves that decrease 
as a function of $\beta$ for fixed $N_{\rm s}$ as discussed earlier.  We only show
the $E_2(\beta, 3)$ curve in Fig.\,\ref{fig: two_project}; its largest value 
is $E(0, 3) =  1.700 $.  The largest values of  $E_2(\beta, N_{\rm s})$ for other
choices of $N_{\rm s}$ are $E_2(0, 5) = 1.744$ 
and $E_2(0, 7) = 1.776$.  Note that for small knowledge breath $\beta$, 
the performance derived from aggregating values via the arithmetic mean 
is larger than the one derived from delegating to a single individual. 
However, beyond a critical $\beta$ value, the reverse is true. This result 
underscores that extending the decision-making process to include input from
all groups is beneficial only if the heterogeneity in their interests and expertise $\beta$
is contained. 

The Delegation method leads to a maximum for $E_{2}(\beta, N_{\rm s})$ as $\beta$ increases. Here,  
the aggregated evaluations $v_i'$ that are used to fill the collective knapsack 
are the ones made by the groups $j_i$ that have the smallest perception error $\sigma_{ij_i} = |t_i - e_{j_i}|$
for each $i=1,2$. Thus, for project $i=1$ one delegates to group $j_1$ and 
for project $i=2$ one delegates to group $j_2$. 
For certain choices of $t_1, t_2$, or for sufficient number
of groups $N_{\rm s}$, it may be that $\beta$ can be set
such that among the expertise levels $e_j$ determined via Eq.\,\eqref{expert} two of them will
coincide with $t_1, t_2$ so that $\sigma_{ij_i} = |t_i - e_{j_i}| = 0$ for $i=1,2$.
These the groups provide exact evaluations, and one delegates to them.
Since there are no errors, the associated $\beta$ leads to the highest likelihood that project $i=2$ of value $v_2 = b > a$ is placed in the collective knapsack,
maximizing $V(t_1, t_2)$.  For large enough $N_{\rm s}$, there may be more than one $\beta$
leading to $\sigma_{ij_i}= 0$,  leading to multiple maxima in $V(t_1, t_2)$. 
The performance $E_2(\beta, N_{\rm s})$, however, is an average over all project types $t_1, t_2$ in
$[t_{\rm min}, t_{\rm max}]$,  hence
one must determine expertise levels $e_j$ 
such that the corresponding $\sigma_{ij}$ are minimized for all possible $t_1, t_2$  project types uniformly distributed within
$[t_{\rm min}, t_{\rm max}]$ and
not just specific, fixed values. The value of $\beta$ that leads to such $e_j$ levels via 
Eq.\,\eqref{expert} is the one that yields the most accurate estimates of the projects, and that maximizes 
$E_2(\beta, N_{\rm s})$. We thus seek to distribute the $e_j$ values in $[t_{\rm min},  t_{\rm max}]$
so that there is adequate expertise to properly evaluate any project type
$t_i$. Mathematically, we seek to ensure that $\sigma_{ij} = |t_i - e_j | $ is smallest for any $t_i$. 
One way to do this is to set $\Delta = t_{\rm max} - t_{\rm min}$ and to define
\begin{align}
\begin{split}
\label{scheme}
e_1 &= t_{\rm min} +  \frac {\Delta}{ 2 N_{\rm s}}, \\
e_j &= e_{j-1} +  \frac {\Delta}{ N_{\rm s}},  \qquad {\mbox {for }}  j \in \{2, \dots, N_{\rm s}\}. \\
\end{split}
\end{align}
This choice ensures that for any $t_i \in [t_{\rm min}, t_{\rm max}]$ 
there exists at least one $j$ such that $ \sigma_{ij} = |t_i - e_{j}| \leq \Delta/ 2 N_{\rm s}.$
This value of $j$ is specifically determined via 
\begin{align}
\begin{split}
\label{scheme2}
{\mbox {If } \,}  (\ell -1)  \frac {\Delta}{ N_{\rm s}} \leq (t_i  -  t_{\rm min})  \leq  \ell \frac {\Delta}{ N_{\rm s}},  \quad {\mbox {then } \, j = \ell }
\quad {\mbox {for }} \, \ell \in \{1, \dots,  N_{\rm s} \}. \\
\end{split}
\end{align}
Coupling the scheme in Eqs.\,\eqref{scheme} with the definition of $\beta$ in Eq.\,\eqref{expert}
we identify $\beta = \beta_{\rm opt}$ that leads to the maximum in $E_2(\beta, N_{\rm s})$ under Delegation as
\begin{equation}
\label{betaopt}
\beta_{\rm opt} (N_{\rm s})= e_{\rm M} - t_{\rm min} - \frac {\Delta}{2 N_{\rm s}}, 
\end{equation}
where we only keep the dependence on $N_{\rm s}$ explicit.  
For the parameters used in Fig.\,\ref{fig: two_project}, 
$e_{\rm M} = 5, \, t_{\rm min} =0,  \, \Delta = 10$,  we find 
$\beta_{\rm opt} (N_{\rm s} = 3) = 10/3$, 
$\beta_{\rm opt} (N_{\rm s} = 5) = 4$, 
$\beta_{\rm opt} (N_{\rm s} = 7) = 30/7$, 
in good agreement with numerical results. 
Furthermore, as $N_{\rm s}$ increases the error $\Delta / 2 N_{\rm s}$ decreases, 
implying less uncertainty in the aggregate evaluations 
$v_i'$. We thus expect $E_2(\beta, N_{\rm s})$ to increase with $N_{\rm s}$, a result that 
 is confirmed in Fig.\,\ref{fig: two_project}. 
As $\beta > \beta_{\rm opt}$ increases, we observe
inflection points in the Delegation curves in Fig.\,\ref{fig: two_project}. 
These emerge as increasing $\beta$ moves some of the expertise levels $e_{j} \neq e_{\rm M}$
outside the $[t_{\rm min}, t_{\rm max}]$ interval so that the corresponding groups are no longer used in the delegation process,
leading to non-trivial changes in the performance. As $\beta$ increases even further,  all expertise levels 
$e_{j} \neq e_{\rm M}$ are outside the $[t_{\rm min}, t_{\rm max}]$ 
interval. In this limit, all evaluations are delegated to the
group with the central expertise $e_{\rm M}$ and   
the Individual and Delegation methods become equivalent.
The threshold of equivalence between the two methods can be estimated
as $\beta_{\rm equiv} (N_{\rm s})=  e_{\rm M} (N_{\rm s} - 1)$. 
For $N_{\rm s} = 3$ we calculate $\beta_{\rm equiv} (3) = 10$, which is
confirmed in Fig.\,\ref{fig: two_project}.  A similar reasoning leads
to conclude that the Individual and Delegation methods
are also equivalent for $\beta = 0$, regardless of $N_{\rm s}$. 

The Median method for $N_{\rm s} = 3$ results in a non-monotonic performance 
$E_{2}(\beta, 3)$,  similar to what observed under Delegation.
The mechanisms driving both trends are similar: for $\beta \to 0$
the limited spread in expertise levels $e_j$ yields relatively large perception errors $|t_i - e_j|$ 
for all groups $j$ and to inaccurate estimates of project values $v_{ij}$. 
Since the resulting median $v_i'$ is not necessarily reflective of actual project values, 
$E_2(\beta, N_{\rm s})$  is relatively low.  As $\beta$ increases,  the likelihood of accurate
evaluations (and of their median values)  increases. The collective knapsack
is now filled with projects whose aggregate values $v_i'$ are closer
to their actual ones, increasing $E_2(\beta, N_{\rm s})$.  However,
to the contrary of the Delegation case where it is enough for any one
of the $N_{\rm s}$ groups to correctly estimate project values,  
here the accurate evaluation must also be the median among the $N_{\rm s}$
evaluations. This extra constraint accounts for a larger $E_2(\beta, N_{\rm s})$ 
and a more easily identifiable $\beta_{\rm opt}$ under 
Delegation than under the Median. As $\beta$ increases further, just as under 
Delegation, $E_2(\beta, N_{\rm s})$  decreases. 

One final observation is that for $\beta \to 0$ all project evaluations
$v_{ij}$ are based on the same perception error $\sigma_{ij} = |t_i - e_{M}|$. Thus,
the more inputs are used to determine the aggregate
value $v_i'$, the more accurate the estimate is. Hence, 
$E(N_{\rm s}, \beta \to 0)$ is largest under  
Arithmetic mean, followed by the Median, Delegation  and the Individual
aggregation methods, respectively.  The latter are equivalent when
$\beta =0$ for all $N_{\rm s}$. 

In the next section, we generalize the above results to 
an arbitrary number of projects $N_{\rm p} > 2$ and introduce extended definitions for 
 $V(t_1, \dots,  t_{N_{\rm p}})$ and $E_{\rm N_{\rm p}}(\beta, N_{\rm s}) $. 
\section{General case with $N_{\rm p}$ projects}
To extend the previous analysis to $N_{\rm p}>2$ projects 
we first define
 $V(t_1, \dots, t_{N_{\rm p}})$ as the expected value 
 of the filled knapsack when there are $N_{\rm p}$ projects to select from. 
Here, the expectation is
taken by averaging over many realizations of the system, each ``replica" carrying its
own  set of $\{v_{ij}\}$ evaluations and its own $\{x_i\}$ list.
Thus, 

\begin{equation}
\label{expfilled}
 V(t_1,\dots, t_{N_{\rm  p}}) =  \mathbb{E} \left[ \sum_{i=1}^{N_{\rm p}} x_i v_i \right]
\end{equation}
 extends the definition of $V(t_1, t_2)$ to  general $N_{\rm p}$. 
 Similarly, we define the performance $E_{\rm N_{\rm p}} (\beta, N_{\rm s} )$ 
as the average of  $ V(t_1,\dots, t_{N_{\rm  p}})$  over all project types. That is, 

\begin{equation} 
E_{\rm N_{\rm p}} (\beta, N_{\rm s} ) \equiv \frac{1}{(t_{\rm max} - t_{\rm min})^{N_{\rm p}}}
 \int_{t_{\rm min}}^{t_{\rm max}} \dots \int_{t_{\rm min}}^{t_{\rm max}} 
 V(t_1,\dots, t_{N_{\rm p}})
\,\mathrm{d}t_1 \dots  \mathrm{d}t_{N_{\rm p}}.  
\label{performance_expectationNp}
\end{equation}
In the previous section, 
we were able to derive analytical results
for $V(t_1, t_2)$ and $E_2(\beta, N_{\rm s})$ in 
Eqs.\,\eqref{vknap} and \eqref{performance_expectation}, respectively, for at least for some aggregation methods.
However, the complexity of the theoretical analysis grows combinatorially 
with $N_{\rm p}$, making 
it very challenging to derive explicit expressions for 
$V(t_1, \dots, t_{N_{\rm p}})$ and $E_{\rm N_{\rm p}} (\beta, N_{\rm s} )$ 
in  Eqs.\,\eqref{expfilled}  and \eqref{performance_expectationNp}
for $N_{\rm p} >2$.
We thus perform numerical simulations 
using the $N_{\rm p} = 2$ analytical results as a reference. 

The first scenario we study includes $N_{\rm p}$ projects 
of integer value $v_i = i$ with 
uniform cost $w_i =1$ for $i \in \{1, \dots, N_{\rm p} \}$.  The project-type interval remains
$[t_{\rm min}, t_{\rm max}]$ and the expertise spread is still as described
in Eq.\,\eqref{expert} with $e_{\rm M} = t_{\rm M}$.  The knapsack budget is 
$W = N_{\rm p}/2$.  Results from this ``baseline" scenario will be compared
to other settings where all quantities remain unchanged except for 
the project cost structure $w_i $  for $i \in \{1, \dots, N_{\rm p}\}$.
The most valuable knapsack in the baseline case ($w_i =1$) is assembled by selecting
the $N_{\rm p}/2$ largest value projects
so that $x_i = 0$ if $i \in \{1, \dots,  N_{\rm p}/2 \}$ and $x_i = 1$ if $ i \in \{N_{\rm p}/2 +1, N_{\rm p} \}$. 
Its value $V_{\rm max} (N_{\rm p})$ is 

\begin{eqnarray}
\label{maxvalue}
V_{\rm max} (N_{\rm p}) = \sum_{i=1}^{N_{\rm p}} x_i v_i  = 
\sum_{i=\frac{N_{\rm p}}{2} + 1}^{N_{\rm p}} i =
\frac{ N_{\rm p} (3 N_{\rm p} +2)}{8}, 
\end{eqnarray}
so that $E_{N_{\rm p}}(\beta, N_{\rm s})  \leq V_{\rm max} (N_{\rm p})$. 
Upon setting $N_{\rm p} = 2$, we recover the scenario discussed in the previous section
for $a=1, b=2$. 

The second scenario we study is that of non-uniform, decreasing costs $w_i =2( N_{\rm p} + 1 -i)/(N_{\rm p} + 1)$ while maintaining 
all other settings unchanged.  This alternative structure for $w_i$ ensures that the combined cost of all projects is 
\begin{equation}
\label{altweight1}
\sum_{i=1}^{N_{\rm p}} w_i = \sum_{i=1}^{N_{\rm p}} 1  = \frac {2}{N_{\rm p} + 1 }\sum_{i=1}^{N_{\rm p}} (N_{\rm p} + 1 -i) = N_{\rm p}, 
\end{equation}
which is the same for uniform costs.  A third interesting scenario is that of increasing costs 
$w_i = 2 i / (N_{\rm p} +1)$ for $i = \{1, \dots, N_{\rm p} \}$ which, 
similarly to Eq.\,\eqref{altweight1}, yield a  combined cost $N_{\rm p}$. 
Here, however, increasing $v_i$ and $w_i$ results in 
quality evaluations $q_{ij}= v_{ij}/w_i$ that are not strongly dependent on $i$. 
Furthermore, our knapsack solver is particularly efficient
when costs and value increase in the same
way, and typically yield the optimal solution even if
the collective values $v_i'$ are different from $v_i$. 
The resulting performances  $E_{N_{\rm p}}(\beta, N_{\rm s})$ are not too dissimilar
across aggregation methods and by construction are much smaller compared to 
the decreasing cost scenario. Hence, we will not further discuss this case. 

In the next section, we numerically evaluate $E_{N_{\rm p}} (\beta, N_{\rm s})$ using different aggregation methods
for $N_{\rm p} = 30$, $W =  N_{\rm p}/2 = 15$ and for $t_{\rm min} =0,  t_{\rm max} = 10$ and $e_{\rm M} = 5$ fixed.
Once all group evaluations  $v_{ij}$ and the corresponding aggregate values $v_i'$ or qualities
$q_i'$ are determined, the collective knapsack is filled. To determine which projects it should include,  we utilize a
classical dynamic programming algorithm specifically designed for the binary knapsack problem 
where values and costs are integer-valued  \citep{martello1987algorithms}; fractional
costs can be included upon scaling by a suitable factor. 
Finally, the performance in Eq.\,\eqref{performance_expectationNp}
is evaluated via Monte-Carlo integration 
based on 500,000 random samplings of the $t_i$ project types
given a specific aggregation method. 

Before discussing our numerical results for uniform and increasing costs, 
 we evaluate the maximal values the knapsack can contain, and the maximal performance. 
 In the uniform case ($w_i = 1$), the largest possible value of the knapsack is $V_{\rm max} (30) = E_{\rm max} (30) = 345$,
and the corresponding cost is the entire budget $W = N_{\rm p}/2 = 15$. 
For decreasing $w_i$, one can show that $V_{\rm max}(30) = E_{\rm max} (30)$ is attained
by including the 21 largest value projects  so that 
$x_i = 0 $ if $i < 10$ and $x_i =1$ if $x_i \geq 10$
resulting in
\begin{eqnarray}
\label{maxvalue2}
V_{\rm max} (30) = \sum_{i=1}^{N_{\rm p}} x_i v_i  = 
\sum_{i=10} ^{30} i = 420. 
\end{eqnarray}
In this case, the cost of the knapsack is
slightly less that the $W=15$ budget since

\begin{equation}
\sum_{i = 10}^{30} w_i = \frac 2 {31}
\sum_{i = 10}^{30}  (31 -i) = \frac{462}{ 31} < 15. 
\end{equation}
Adding the next best value project, with $i = 9$, would augment
the knapsack cost by 44/31 increasing the total cost to $506/31 > 15$,
thus exceeding the $W =15$ budget.  Finally, for increasing
$w_i$ the maximal value  $V_{\rm max}(30) = E_{\rm max} (30)= 232$ 
and is attained by including the 20 lowest value projects and the 22$^{\rm nd}$ so that 
$x_i = 1$ if $i \leq 20$,  $x_{22} = 1$ and $x_i = 0$ otherwise. In this case, 
$V_{\rm max}(30) = 232$ and the total cost is $14.97$. 

\subsection{Numerical results for $N_{\rm p} = 30$ projects and various aggregation methods}

\begin{figure}
	\centering
	\includegraphics{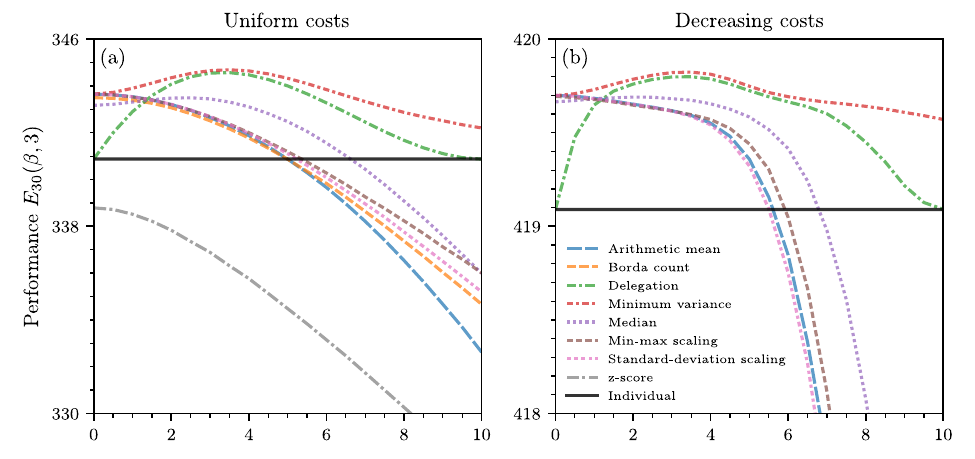}
		\includegraphics{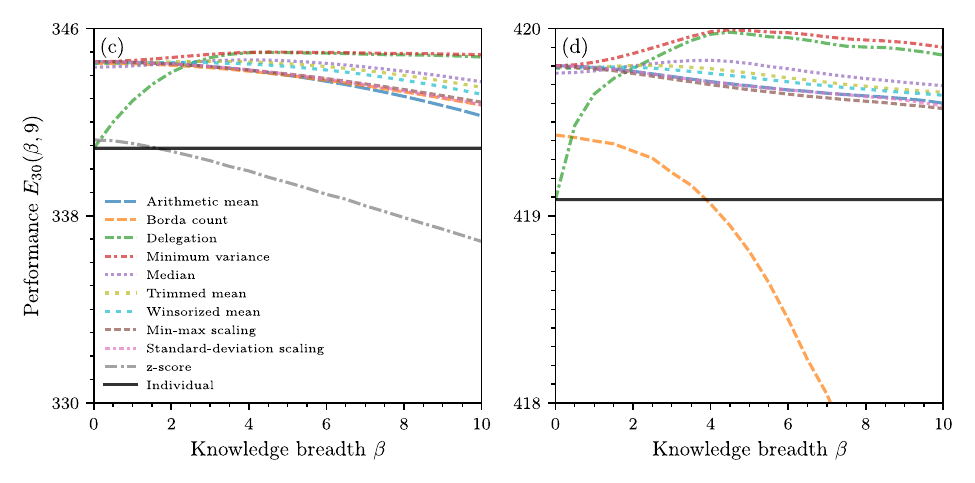}
        \captionsetup{justification=justified}
\caption{Performance $E_{30}(\beta, N_{\rm s})$ as defined in 
Eq.\,\eqref{performance_expectationNp} 
for $N_{\rm p} =30$ projects of value $v_i= i$, for $i \in  \{1, \dots,  N_{\rm p} \}$ for various aggregation methods.
In panels (a,b) we plot $E_{\rm 30}(\beta, 3)$; 
in panels (c,d) we plot $E_{\rm 30}(\beta, 9)$.
Costs are uniform ($w_i = 1$) in panels (a,c) and decreasing 
($w_i =  2 (N_{\rm p} +1 -i)/ (N_{\rm p} + 1)$) in panels (b,d).
Group expertises are within $[e_{\rm M} - \beta, e_{\rm M} + \beta]$ where $e_{\rm M} = 5$ and $\beta$
can vary. Project types are between $[t_{\rm min} , t_{\rm max}]$ where $t_{\rm min}=0, t_{\rm max} =10$ and the collective knapsack budget is set at
$W=N_{\rm p}/2=15$. For the chosen parameters, 
$ E_{\rm max} (30) =345$ in panels (a,c) (uniform costs)  and
$ E_{\rm max} (30) =420$ in panels (b,d) (decreasing costs). 
We numerically surveyed all aggregation methods described in the text. Results from the Yes-or-no voting method do not appear in
any of the panels as the resulting performances are below the chosen vertical axes scales. Relevant ranges for this method are: 
(a): 245 -- 269;  (b): 217 -- 275;  (c): 255 -- 296; (d): 239 -- 352.  
Similarly, Borda count results do not appear in panel
(b), the relevant range is 404 -- 417; $z$-score results do not appear in panels (b,d),  the relevant ranges are
(b): 189 -- 191; (d): 189 -- 191.  
For $N_{\rm s} = 3$ in panels (a,b)  the Trimmed mean and the Winsorized mean are equivalent to the Median. 
The aggregation method that results in the largest performance $E_{\rm 30}(\beta, N_{\rm s})$ in all panels
is the Minimum variance, closely followed by Delegation.
Both methods bias the evaluation of projects in favor of groups whose expertise is closest to the project type. 
For all aggregation methods and fixed $\beta, N_{\rm s}$, the performance
$E_{\rm 30}(\beta, N_{\rm s})$ is largest for decreasing cost structures. 
Increasing $N_{\rm s}$ improves the performance in all panels;  increasing $\beta$ is generally detrimental 
except for select aggregation methods where intermediate values
of $\beta$ yield a maximum performance. These methods are
the Minimum variance, Delegation, and the Median for $N_{\rm s} =3$
where a larger spread $\beta$ ensures that at least one group's expertise 
is close to all possible project types.}
 \label{fig:many_projects}
\end{figure}

Here, we present  results from numerical studies of collective knapsacks 
constructed using various aggregation methods
for $N_{\rm p} =30$,  $W = N_{\rm p}/2 =15 $, $t_{\rm min} =0,  t_{\rm max} = 10$ and $e_{\rm M} = 5$. 
When using the asymmetric estimators
 (the Trimmed and Winsorized means) we set $\alpha=0.2$, and set $p$
 to be the value of $\alpha N_{\rm s}$ rounded to nearest integer.
Some of our results are shown in Figs.\,\ref{fig:many_projects}(a,c), 
corresponding to the uniform cost case $w_i = 1$ for $N_{\rm s} =3$ and
 $N_{\rm s} =9$ respectively, and in
Figs.\,\ref{fig:many_projects}(b,d), corresponding to the decreasing cost 
case $w_i =   2(N_{ \rm p} + 1 -i) /(N_{\rm p} + 1) $  for $N_{\rm s} =3$ and
 $N_{\rm s} =9$, respectively.   More detailed plots
 are in the Appendix, where we show the performance $E_{\rm 30} (\beta, N_{\rm s})$ for each of the twelve aggregation
 methods for several values of $N_{\rm s}$ using the same
$N_{\rm p} =30$,  $W = N_{\rm p}/2 =15 $, $t_{\rm min} =0,  t_{\rm max} = 10$ and $e_{\rm M} = 5$
parameters.
  
The most noticeable feature we observe is 
that, for fixed $N_{\rm s}$, weight distributions strongly affect the performance and that 
the decreasing cost structure leads to collective knapsacks with 
higher overall value. This trend emerges from having associated
large-value projects with low costs, making their inclusion in the knapsack very advantageous.
This trend emerges for all values of $N_{\rm s}$ that we probed. 
Another observation is that among all aggregation 
methods we surveyed, regardless of cost structure and number of groups
$N_{\rm s}$,  the one that yields the largest performance is
the Minimum variance, closely followed by Delegation. Both of these methods
prioritize the evaluation made by groups whose expertise is closest to project type. 

We also find that 
for most aggregation methods $E_{\rm N_{\rm p}}(\beta, N_{\rm s})$ increases with $N_{\rm s}$  for fixed $\beta$,
for both uniform and decreasing costs. Here, increasing the number of stakeholder groups implies a more numerous, and more 
diverse, set of expertises, smaller perception errors $\sigma_{ij}$, more accurate
aggregated estimates,  and thus larger performances. 
This is observed by comparing the $N_{\rm s} = 3$ curves in 
Fig.\,\ref{fig:many_projects}(a) with the $N_{\rm s} = 9$ ones in Fig.\,\ref{fig:many_projects}(c) 
and similarly by comparing their counterparts in 
Figs.\,\ref{fig:many_projects}(b,d). 
The only aggregation method for which $E_{\rm N_{\rm p}}(\beta, N_{\rm s})$
is independent of $N_{\rm s}$ is the Individual one, since here 
the only evaluations used are those made by the group with central expertise $e_{\rm M} = 5$ regardless
of $N_{\rm s}$.  When $\beta =0$ or $\beta \to \infty$, 
the performance  under Delegation is
the same as under the Individual aggregation method.
This is also observed in Fig.\,\ref{fig: two_project}
for $N_{\rm p} =2$. 

Once  $N_{\rm s}$ is fixed, and for small to moderate values of $\beta$, 
most aggregation methods, direct or indirect, achieve better performance
compared to the Individual method, with few exceptions. One of them is the
$z$-score. This is to be expected as under $z$-score about half the projects are 
assigned a negative quality and discarded, as their inclusion in the collective knapsack
would decrease the overall value. Hence, when using
$z$-scores, the knapsack is sub-optimally populated and there is 
an excess available budget.  Another method that performs
poorly is Yes-or-no voting, due to the emergence of many ties among 
for projects of the same quality that hinder selection of the highest valued. 
Although the $E_{\rm N_{\rm p}}(\beta, N_{\rm s})$ curves for the Yes-or-no voting method 
are not included in Fig.\,\ref{fig:many_projects} due to their low values compared
to the chosen scale, the Yes-or-no voting aggregation method 
is the only one whose performance increases with $\beta$ for the given parameters. 
In this case,  increasing $\beta$ leads to more heterogeneity in project evaluations, 
decreasing the number of ties and thus yielding a larger performance. 

For $N_{\rm s} =3$, the inclusion of indirect aggregation methods that cannot be easily evaluated 
through analytical methods (such as the Borda count or the Min-max scaling)
results in additional performance curves $E_{30}(\beta, 3)$ that decrease substantially as
$\beta$ increases compared to the
ones obtained for direct methods as shown in 
Figs.\,\ref{fig:many_projects}(a,b). Indirect methods are competitive with direct ones
only when the expertise spread $\beta$ is contained. Increasing the number of
stakeholder groups to $N_{\rm s} =9$ results in larger performances
for all methods, particularly for the indirect ones and large $\beta$. 
$E_{30}(\beta, 9)$ is still larger for direct
methods than for indirect ones, but the gap between them is reduced
compared to the $N_{\rm s} =3$ case. 

Many findings obtained by setting $N_{\rm p}=2$ in the uniform
cost case are robust to increasing the number of projects to
$N_{\rm p} =30$ as can be seen from comparing 
Fig.\,\ref{fig: two_project} with Figs.\,\ref{fig:many_projects}(a,c). In particular, curves corresponding
to the Arithmetic mean decrease in $\beta$, whereas those obtained under 
Delegation display a maximum.  The value $\beta_{\rm opt}$ for which 
the performance is maximized under Delegation for $N_{\rm p} =30$ is the same as for 
$N_{\rm p} =2$  and is given in Eq.\,\eqref{betaopt}. Since computing the performance
involves averaging over all project types, its maximum is found 
by equally spacing the group expertise levels on the range of the project types, regardless of the
actual number of available projects.  Similarly, the Median aggregation method
leads to a shallow maximum.  There are, however, no inflection points. 
Furthermore, increasing the number of groups $N_{\rm s}$ typically increases the
performance for fixed $\beta$. 

Another interesting feature of Fig.\,\ref{fig:many_projects} is that, as predicted, the $z$-score aggregation method,
a very natural protocol due its ubiquity in statistics, 
performs poorly for both uniform and decreasing cost structures, and for both $N_{\rm s} =3, 9$. This is because, 
by construction, each group assigns a negative  $z$-score to about half the projects thorough Eq.\,\eqref{zscore}; 
these are the ones whose value is less than the mean. 
Thus, there is a high likelihood that the aggregated score $q_i'$, given by the sum of the $z$-scores presented by all groups
as per Eq.\,\eqref{zscoreavg} is also negative. Projects with a negative evaluation are not included
in the collective knapsack as they add cost without increasing the value. 
As a result, under $z$-score it is very likely that the collective knapsack will contain 
a suboptimal number of projects, leaving a residual budget that cannot be utilized.  
We corrected for this feature of the $z$-score
by introducing a variant aggregation method, 
Standard-deviation scaling, which results in positive
aggregated scores $q_i'$ for all projects. 
As can be seen in
Fig.\,\ref{fig:many_projects}, this protocol outperforms the $z$-score in all cases. 
Other variants of the $z$-score can also be devised, including greedy methods. 

Overall, our results imply that the best overall performance
is obtained when decision-making is based on the input of 
multiple groups whose expertise carries a moderate spread
(large $N_{\rm s}$, intermediate $\beta$) 
and that direct methods perform better than indirect ones. 
For the parameters used in this paper, the best aggregation method is 
Minimum variance, where all groups partake in the decision-making process but 
the input of those whose expertise is most closely aligned with project type
is favored.  Good performances also arise under Delegation. 

\section{Information errors}

As shown in Fig.\,\ref{fig:many_projects}, the Minimum variance and
Delegation methods yield the best performances. However, in both
cases, the central decision-maker must compare expertises $e_j$ of all
the $N_{\rm s}$ groups to determine which are associated with the
smallest (or most relevant) perception errors
$\sigma_{ij}$. Alternatively, it is the groups themselves that
disclose this information and make comparisons amongst themselves to
identify which of them has the most suitable $\sigma_{ij}$.

The need to compare all $\sigma_{ij}$ values implies that all project
expertises $e_j$ are known. Here, we assume that there may be
information errors so that either the central decision-maker, or the
$N_{\rm s}$ groups, are not able to properly assess the $e_j$
expertise levels.  Project types $t_i$ are assumed to be known for all
$i \in \{1, \dots, N_{\rm p}\}$.  We thus introduce the probability
$r$ ($0 \leq r \leq 1$) that the aggregation process is flawed.

Under Delegation, $r$ is the probability that project estimation is
delegated to a randomly selected group rather than to the one whose
expertise is most closely aligned with the project type. The limit $r
= 0$ implies that the delegation process is always error-free and in
favor of the group with the smallest $\sigma_{ij}$.  If $r = 1$,
instead, each group has the same probability of being selected,
regardless of its expertise level.  In practice, once $r$ is given, we
delegate the decision with probability $r / N_{\rm s}$ to any of the
$N_{\rm s} - 1$ non-optimal groups, and with probability $1 - (N_{\rm
  s} - 1) r / N_{\rm s}$ to the optimal group $k^*$ as per
Eq.\,\eqref{delegate}.  Under Minimum variance, $r$ is the probability
that project estimation is achieved via the Arithmetic mean rather
that through optimally weighting evaluations through
Eqs. \eqref{linear_combination} and \eqref{linear_combination2}.  The
limit $r = 0$ implies that Minimum variance is always performed
correctly whereas $r = 1$ results in the Arithmetic mean being used in
all evaluations.  Since $r$ is the likelihood that a project is
evaluated incorrectly, the number of projects that are incorrectly
evaluated follows a binomial distribution with expectation $r N_{\rm
  p}$.
 
\begin{figure}
    \centering
    \includegraphics{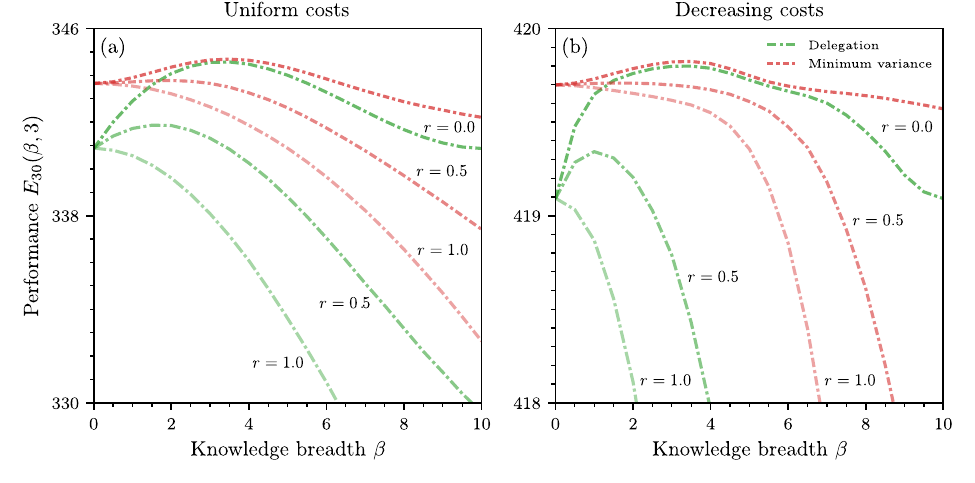}
        \captionsetup{justification=justified}
    \caption{Performance $E_{\rm 30} (\beta,3)$ under Delegation
      (green curves) and Minimum variance (red curves) subject to
      different probabilities of information error $r = 0, 0.5, 1$ for
      uniform (panel (a), $w_i=1 $) and decreasing costs (panel (b),
      $w_i = 2(N_{\rm p} +1 -i )/(N_{\rm p} + 1)$).  Group expertises
      are within $[e_{\rm M} - \beta, e_{\rm M} + \beta]$ where
      $e_{\rm M} = 5$ and $\beta$ can vary. Project types are between
      $[t_{\rm min}, t_{\rm max}]$ where $t_{\rm min} = 0, t_{\rm max}
      = 10$ and the collective knapsack budget is set at $W = N_{\rm
        p}/2 = 15. $ The error-free case $r=0$ (top-most green and red
      curves) yields the best performance for both aggregation
      methods; the full-error case $r=1$ (lowest green and red curves)
      results in the worst performance for both. Increasing the
      expertise heterogeneity $\beta$ is particularly detrimental
      under Delegation, since decisions are made by one group only,
      and this group may completely misevaluate the project at
      hand. When $r=0$, the Minimum variance method reduces to the
      Arithmetic mean whereby averaging the value of projects using
      $N_{\rm s}$ groups yields a less inaccurate evaluation compared
      to when a single, random group is queried.  }
    \label{fig:delegation_error}
\end{figure}

In Fig.\,\ref{fig:delegation_error}, we plot $E_{30}(\beta, N_{\rm
  s})$ for $N_{\rm s}=3$ under Delegation and Minimum variance subject
to different levels of information error $r \in \{0, 0.5, 1\}$ for
uniform and decreasing cost structures.  As $r$ increases, performance
levels decline for both aggregation methods across both cost
structures, with the effect being particularly pronounced under
Delegation. When $r=0.5$, the optimal group in Delegation is selected
with a probability of $1-2 \times 0.5/3=2/3$, yet performance still
drops significantly. This decline is even more pronounced in the case
of decreasing costs, as illustrated in
Fig.~\ref{fig:delegation_error}(b). The limit $r = 1$ represents the
scenario where selection of the most suitable group is completely
random. Thus, increasing the expertise heterogeneity $\beta$ is
detrimental since all evaluations have the same likelihood of being
selected.  Minimum variance is more robust to increasing the
information error to $r = 1$ since this case reduces to the Arithmetic
mean, whereby the input of all groups is used, tempering the effect of
any unrealistic evaluation.

\section{The Borda count}

One interesting observation from Fig.\,\ref{fig:many_projects} is that
the performance of the Borda count method is on par with several
others and actually even exceeds that of the Arithmetic mean in the
uniform cost case in panels (a,c).  However, its performance is much
worse performance than that of all other methods in the decreasing
cost case in panels (b,d).  Why is this the case?

Unlike direct methods that are value-based, the Borda count scores
projects on a pre-determined range based on project qualities
resulting in the minimum score being set at zero and the maximum score
being capped at $N_{\rm p} - 1$, regardless of the actual quality
values and associated spread.  As pointed out by
\cite{boettcher2024selection}, capping scores is an effective way to
include outliers in the evaluations and can more accurately identify
the ``best" projects especially when the uncertainty in evaluating
projects is very large.  Thus, in the uniform cost case, where quality
and value are proportional (or the same upon setting $w_i =1$ for all
$i \in \{1, \dots, N_{\rm p} \}$), and where cost constraints only
impose an upper limit on the number of projects that can be selected,
the Borda count performs well, even when the expertise spread is
large.

However, when weights are not uniformly assigned, the interplay
between quality, cost, and the structure of the Borda scores becomes
more relevant.  In some cases, projects with the highest quality are
not included in the collective knapsack because their Borda scores do
not lead to maximal knapsack total value.  This is due to the mapping
of the Borda score onto a finite interval that ``smooths out"
outliers.  For the sake of argument, consider a simple example with no
perception error $\sigma_{ij} =0$ and where three possible projects
can be used to fill a collective knapsack of weight $W = 1$.  We set
values and cost, and derive the corresponding quality, Borda counts,
and collective knapsack argument, as follows

\begin{equation}
\begin{aligned}
v_1 &= 10, & w_1 &= 0.1, & q_1 &= 100,  \qquad &s_1 = q'_1 & = 2,  \qquad & q'_1 w_1 & =  0.2; \\
v_2 & = 2, & w_2  &= 1, & q_2 & = 2,  \qquad &s_2 =   q'_2 & = 1,  \qquad & q'_2 w_2 & = 1; \\
v_3 & = 1, & w_3  & = 0.9, & q_3 & = 1.11, \qquad & s_3  = q'_3 &= 0,  \qquad & q'_3 w_3 & = 0. 
\end{aligned} 
\end{equation}
Since there is no error, all groups will evaluate projects in the same way.
Hence, the goal is to maximize 

\begin{equation}
    \max \sum_{i=1}^{N_{\rm p}} q'_i w_ix_i, 
    \label{max20}
\end{equation}
and use the $x_i$ values to 
determine the $\beta$-independent performance $ E_{3} $ given by 
\begin{equation}
\label{expfilled2}
E_{3} = 
\sum_{i=1}^{N_{\rm p}} x_i v_i, 
\end{equation}
where Eq.\,\eqref{expfilled2} is the same as Eq.\,\eqref{expfilled}
under the assumption that $\sigma_{ij} =0$ so that the performance is
independent of expertise and project type and on the number of
stakeholder groups.  In this example, project 1 is associated with the
highest quality and $E_{3}$ is optimized by always selecting projects
1 and 3, which carry a total cost $w_1 + w_3 = W = 1$, and yield the
performance $v_1 + v_3 = 11$.  Selecting project 2 results in a
collective knapsack of cost $w_2 = W = 1$ and of performance $v_2 =
2$, which is less than the valued obtained by selecting projects 1 and
3.

When using the Borda count, the collective knapsack in
Eq.\,\eqref{max20} is maximized by including only project 2, since
$q'_2 w_2 = 1$ is greater than the combined contribution of projects 1
and 3, for which $q'_1 w_1 + q'_3 w_3 = 0.2$.  This outcome is due to
the fact that although the quality discrepancy between projects 1 and
2 is very large $(q_1 = 100 > q_2 =2)$ the Borda count reduces this
gap to just one unit $(q'_1 = 2 > q'_2 =1)$. Furthermore, when filling
the collective knapsack, due to the non-uniform cost structure, the
contribution of project 2 exceeds that of project 1 $(q'_1 w_1= 0.2 <
q'_2 w_2 =1)$.  If costs were uniform and $w_1 = w_2$ this ``switch"
could not occur.  This example is illustrative of why the Borda count
is not expected to perform well in the decreasing cost scenario as
observed in Fig.\,\ref{fig:many_projects}.  Here, the presence of
perception errors in evaluating projects and the decreasing cost
structure will facilitate ``switches" similar to the one described
above leading to sub-optimal projects being added to the collective
knapsack.

\begin{figure}
    \centering
    \includegraphics{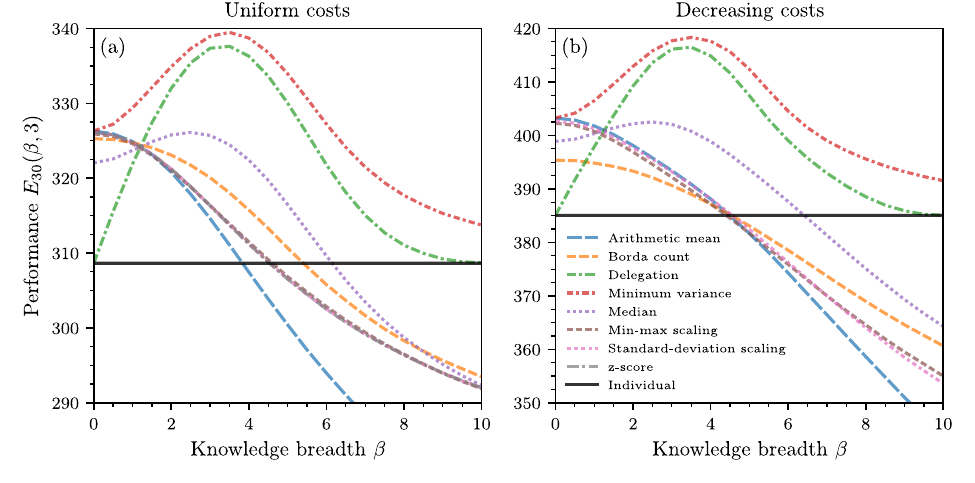}
    \captionsetup{justification=justified}
    \caption{Performance $E_{30}(\beta, 3)$ as defined in Eq.\,\eqref{performance_expectationNp}; 
    in panel (a) we used uniform costs and in panel (b) we used decreasing costs. 
    Other settings and parameters are the same as in Fig.\,\ref{fig:many_projects} except here
    we quadruple the perception error to $\sigma_{ij} = 4 |e_i - t_j|$.
    As can be seen, most performances decrease relative to the ones observed in Fig.\,\ref{fig:many_projects}, 
    however the performance under Borda count improves dramatically, especially in the decreasing cost case,
    outperforming other methods such as Min-max scaling,  Arithmetic mean and Standard deviation scaling
    for large enough $\beta$. Minimum variance and Delegation remain the best performing methods and their maxima
    is located at the same $\beta_{\rm opt}$ value as in Fig.\,\ref{fig:many_projects}. Results derived under the Trimmed and Windsorized
    means coincide with those derived under the Median; results from the Yes-or-no voting method do not appear in panels
    (a,b) since the resulting performances are below the chosen vertical axes scales. Relevant ranges for this method are: 
   (a): 267 -- 277; (b) 277 -- 310. 
}
\label{fig:noise_panel}
\end{figure}

The Borda count is instead very useful when outliers are present as it
imposes a preset range of values that tempers their effect. This is
true of all rescaled methods.  When aggregating values using the
Arithmetic mean, for example, outlier groups may assign an anomalously
high or low value to a specific project due to large perception errors
skewing the aggregate result. However, under the Borda count or other
rescaled methods, the impact of outliers is mitigated due to the fixed
range.  This can be seen in Fig.\,\ref{fig:noise_panel} where we
quadruple the perception error and set it to $\sigma_{ij} = 4 |e_i -
t_j|$ for $N_{\rm s} =3$ stakeholder groups.  All other parameters and
settings are the same as in Fig.\,\ref{fig:many_projects}(a,b).  Upon
comparing Figs.\,\ref{fig:many_projects} and \ref{fig:noise_panel},
one can observe that in panels (a), for uniform costs, all
performances decrease when quadrupling the perception error but the
least reduction is associated to the Borda count. This method
outperforms, albeit marginally, many other methods for large $\beta$.
The effectiveness of the Borda count when increasing the perception
error is much more pronounced in the case of decreasing costs, as can
be seen by comparing panels (b) of Figs.\,\ref{fig:many_projects} and
\ref{fig:noise_panel}.  In Fig.\,\ref{fig:many_projects}(b) results
from the Borda count do not appear as they are lower than the chosen
vertical axis scale. Here, the Borda count yields the worst
performance of all methods for all $\beta$. In
Fig.\,\ref{fig:noise_panel}(b), instead, when the perception error is
larger, the performance of Borda count visibly improves, even
surpassing that of the Arithmetic mean, Min-max scaling, and Standard
deviation scaling for large $\beta$. Delegation and Minimum variance,
however, remain the best performing methods.

\section{Discussion and conclusion}
In this paper, we developed a mathematical model of
collective-decision-making where the input of several stakeholder
groups, each with their own priorities and expertises, are taken into
account. We considered the general scenario of tactical
decision-making whereby a set of projects that aim to achieve the same
strategic objective is given. A subset of them must be chosen with the
objective of maximizing the total long-term benefit, or value, under
cost constraints.  We used the perception noise scenario where,
although projects have an intrinsic value, each group carries out its
own evaluation, based on specific priorities or expertise, leading to
heterogenous assessments. For each project, it is thus necessary to
aggregate the diverse group estimates and derive a unique evaluation
of its benefits.  This final aggregated evaluation may, or may not, be
aligned with the project's intrinsic value.  All aggregation methods
are then combined with knapsack solvers so that a ``collective"
knapsack is filled; the selected projects represent the ``portfolio"
of projects to be implemented.

We proposed different protocols to aggregate group evaluations,
distinguishing between direct and indirect methods. The direct ones
involve using group evaluations as they are, the indirect ones map
existing group evaluations onto scores or rescaled quantities on
predetermined intervals. We also classify these methods as private or
transparent, depending on how much information about the groups must
be made public in order for the aggregation process to be carried out,
and on whether they are based on representative or direct voting.  The
efficiency of each method is determined via a performance metric that
encapsulates the overall value of the final subset of selected
projects.

While some of the aggregation methods studied in this paper have been
already introduced in the literature as direct (Arithmetic mean,
Median, Individual, Delegation) and indirect protocols (Yes-or-no
voting, Borda count), several others are novel to this work. In
particular, we expand the repertoire of direct methods by adding the
Trimmed and Winsorized means, as well as the Minimum-variance
method. Additionally, we add to the repertoire of indirect methods by
including the Min-max, $z$-score, and Standard-deviation scaling
techniques.

Previous research in portfolio selection was limited to scenarios with
uniform project costs~\citep{boettcher2024selection}; in contrast,
this work addresses the more realistic case of heterogeneous
costs. Specifically, we compare and contrast results arising from
assigning projects uniform costs to those arising from assigning lower
costs to projects with larger intrinsic value.  Including a novel,
non-trivial cost structure modifies the performance of several
aggregation methods; most strikingly when using the Borda count. This
method performs relatively well when costs are uniform but is less
effective when costs decrease, particularly in cases where there is a
wide variation in expertise. Here, the performance of the Borda count
is inferior to that of many other methods due to the interplay between
cost structure, perception error, and mapping qualities onto
predefined ranges (\ie, through the Borda scores).

Many of the new aggregation methods proposed in this work outperform
the existing ones, even in the case of uniform costs.  For almost all
aggregation methods, and regardless of cost structure, we find that
the performance improves as the number of stakeholder groups increases
and as the heterogeneity in their expertise levels decreases.  Among
the exceptions to the above trends are the best performing aggregation
methods, Minimum variance and Delegation where instead the performance
reaches a maximum at an optimal value of the expertise spread.
Specifically, Minimum variance and Delegation, perform particularly
well when there is enough, but not excessive, diversity in the
expertise range so that at least one of the groups to properly
evaluate any project. Both Minimum variance and Delegation rely on
groups sharing their priorities or expertise levels with one another
or with the central decision-maker.  In some cases, especially for low
project and stakeholder group numbers, we were also able to provide
analytical insight into the performance of several aggregation
methods, advancing the state-of-the-art since previous analytical
results were restricted to deriving the maximum performance for the
existing methods as an upper bound.

A natural question arises: of the twelve aggregation methods
introduced in this paper, each with their own performance,
transparency and degree of representativeness, which one should be
chosen when collective decisions must be made? Within the context of
our work, if expert groups can be identified for each project and
society as a whole is cohesive enough to permit biases in favor of
their judgements, then our results suggest that Minimum variance
should be prioritized since it performs the best in all scenarios
analyzed. Even when expertise levels cannot be accurately identified,
the performance of Minimum variance is still bounded from below by the
Arithmetic Mean, which is itself an effective aggregation
method. Delegation also performs well, but it requires even greater
societal trust, as each project is evaluated solely by the group with
the highest expertise, while the input from all other groups is
disregarded. However, this method is only effective when expert groups
can be identified with near certainty. If experts cannot be
identified, and the $\sigma_{ij}$ values are unknown, then Median and
Arithmetic mean provide effective rules.  Which of the two should be
utilized depends on the spread of expertise $\beta$ among constituent
groups.  Finally, the Borda count is useful when value proxies must be
used, although its performance is subpar for decreasing cost
structures and for large heterogeneity in group expertise spread.

The conclusions above pertain to the specific parameter settings and
scenarios studied in this work and may not extend to other
realizations.  For example, although we performed extensive numerical
studies to study how each aggregation method performs upon varying the
expertise spread $\beta$ and the number of stakeholder groups $N_{\rm
  s}$, we kept the value of the budget fixed at $W = N_{\rm p}/2$ and
considered only two cases for $N_{\rm p}$, namely $N_{\rm p} = 2$ and
$N_{\rm p} = 30$. Further investigations would include varying $W$ and
$N_{\rm p}$, the relationship between them, or introducing a
relationship between $W$ and $N_{\rm s}$, representing the scenario
whereby a society with a larger number of constituent groups is
warranted a larger budget.

Many of the parameters chosen in this work are fixed and derived from
the existing literature \citep{boettcher2024selection,
  csaszar2013organizational} including the width of the project-type
interval, the center of the expertise interval, the intrinsic project
values. Similarly, the definition of the perception error when
evaluating projects as the discrepancy between expertise and project
type is taken from the literature on organizational decision-making in
management science.  We chose to embed our work within this context so
that our novel aggregation methods, analytical results, and findings
can be used as natural extensions to the existing literature.
However, all constructs described above could be modified to include
qualitatively different scenarios, for example by imposing that the
center of the expertise and project-type interval be different,
allowing the expertise level to be randomly distributed instead of
ordering them according to Eq.\,\eqref{expert}, or by introducing
alternative ways of defining the perception errors $\sigma_{ij}$. Another interesting extension could involve studying hierarchical aggregation structures~\citep{bottcher2022examining}. Additionally, as discussed earlier in this article, one might consider going beyond perception noise $\epsilon_{ij}^{\rm p}$ and assume that projects have tangible impacts on each group, either detrimental or advantageous. In this approach, the intrinsic value of project $i$ would not be $v_i$, but rather its aggregated counterpart $v_i'$, reflecting the tangible consequences of the project on all groups.

\section{Acknowledgements}

This work was supported by the Army Research Office through
W911NF-23-1-0129 (YG, LB, MRD) and by the National Science Foundation
through grant MRI-2320846 (MRD).  LB also acknowledges funding from
hessian.AI and thanks Ronald Klingebiel for useful discussions.
We thank the Santa Fe Institute for organizing the 
Collective Intelligence Symposium where the authors held their first discussions on this work.



\section{Competing interests}

The authors declare that they have no known competing financial
interests or personal relationships that could have appeared to
influence the work reported in this paper.

\section {Appendix} 
\begin{figure}[t!]
    \centering
    \includegraphics[width=\linewidth]{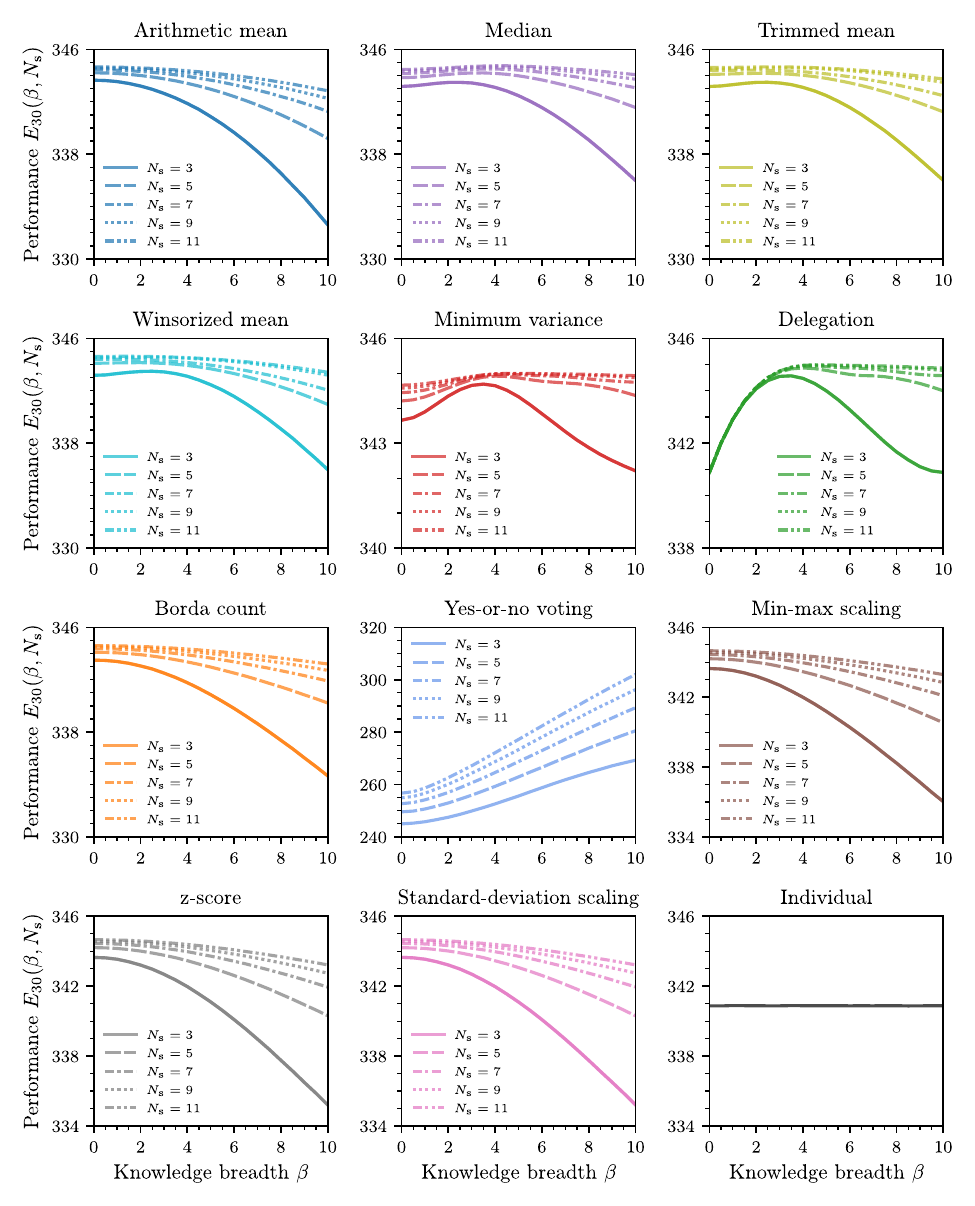}
    \captionsetup{justification=justified}
    \caption{Performance $E_{\rm 30} (\beta, N_{\rm s})$ as defined in
      Eq.\,\eqref{performance_expectationNp} and upon aggregating
      projects using all twelve methods surveyed in this paper. We set
      $N_{\rm s} = 3,5,7,9,11$ and $v_i = i$ and assume uniform costs
      $w_i =1$. All other parameters are the same as in
      Fig.\,\ref{fig:many_projects}: $N_{\rm p} =30$ , $W= N_{\rm
        p}/2=15$, $t_{\rm min} =0, t_{\rm max} =10$ and $e_{\rm M}
      =5$.}
    \label{fig:appendix1}
\end{figure}

\begin{figure}[t!]
    \centering
    \includegraphics[width=\linewidth]{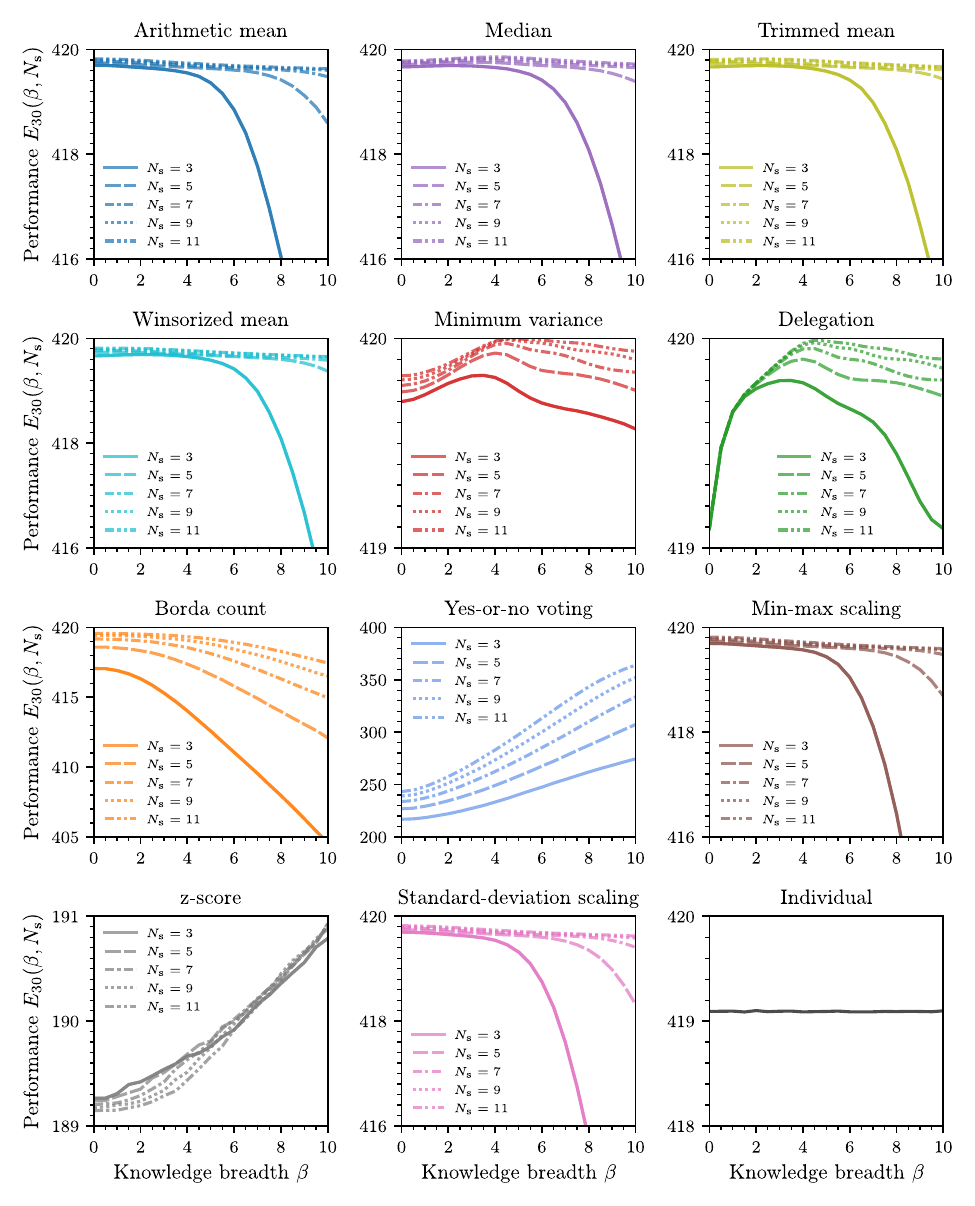}
    \captionsetup{justification=justified}
    \caption{Performance $E_{\rm 30} (\beta, N_{\rm s})$ as defined in
      Eq.\,\eqref{performance_expectationNp} and upon aggregating
      projects using all twelve methods surveyed in this paper. We set
      $N_{\rm s} = 3,5,7,9,11$ and $v_i =i$ and assume decreasing
      costs $w_i =2( N_{\rm p} + 1 -i)/(N_{\rm p} + 1) $. All other
      parameters are the same as in Fig.\,\ref{fig:many_projects}:
      $N_{\rm p} =30$ , $W= N_{\rm p}/2=15$, $t_{\rm min} =0, t_{\rm
        max} =10$ and $e_{\rm M} =5$.}
    \label{fig:appendix2}
\end{figure}

For completeness, we show the performance $E_{30}(\beta,N_{\rm s})$
for each of the twelve aggregation methods for several values of
$N_{\rm s}$ using the same $N_{\rm p} =30$ , $W= N_{\rm p}/2=15$,
$t_{\rm min} =0, t_{\rm max} =10$ and $e_{\rm M} =5$ parameters used
for Fig.\,\ref{fig:many_projects}.  Results for the uniform cost case
($v_i =i, w_i =1$) are shown in Fig.\,\ref{fig:appendix1}; results for
the decreasing cost case ($v_i = i, w_i =2( N_{\rm p} + 1 -i)/(N_{\rm
  p} + 1)$) are shown in Fig.\,\ref{fig:appendix2}.

\newpage

\bibliographystyle{SageH}
\bibliography{main.bib}

\begin{thebibliography}{44}
\providecommand{\natexlab}[1]{#1}
\providecommand{\url}[1]{\texttt{#1}}
\providecommand{\urlprefix}{URL }
\expandafter\ifx\csname urlstyle\endcsname\relax
  \providecommand{\doi}[1]{DOI:\discretionary{}{}{}#1}\else
  \providecommand{\doi}{DOI:\discretionary{}{}{}\begingroup
  \urlstyle{rm}\Url}\fi

\bibitem[{Arribillaga and Bergantinos(2022)}]{Arribillaga2022}
Arribillaga RP and Bergantinos G (2022) Cooperative and axiomatic approaches to
  the knapsack allocation problem.
\newblock \emph{Annals of Operations Reseach} 318: 805--830.

\bibitem[{Aziz and Shah(2021)}]{Aziz2021}
Aziz H and Shah N (2021) Participatory budgeting: Models and approaches.
\newblock In: Rudas T and P{\'e}li G (eds.) \emph{Pathways Between Social
  Science and Computational Social Science: Theories, Methods, and
  Interpretations}. Cham, Switzerland: Springer International Publishing, pp.
  215--236.

\bibitem[{Baum(2020)}]{Baum2020}
Baum SD (2020) Social choice ethics in artificial intelligence.
\newblock \emph{AI $\&$ Society} 35: 165--176.

\bibitem[{Bell et~al.(1988)Bell, Raiffa and Tversky}]{Bell1988}
Bell DE, Raiffa H and Tversky A (eds.)  (1988) \emph{Decision making:
  {D}escriptive, normative, and prescriptive interactions}.
\newblock Cambridge, UK: Cambridge University Press.

\bibitem[{Benade et~al.(2021)Benade, Nath, Procaccia and
  Shah}]{benade2021preference}
Benade G, Nath S, Procaccia AD and Shah N (2021) Preference elicitation for
  participatory budgeting.
\newblock \emph{Management Science} 67: 2813--2827.

\bibitem[{Bhagat et~al.(2014)Bhagat, Kim, Muthukrishnan and
  Weinsberg}]{Bhagat2014}
Bhagat S, Kim A, Muthukrishnan S and Weinsberg U (2014) The {Shapley} value in
  knapsack budgeted games.
\newblock In: Liu TY, Qi Q and Ye Y (eds.) \emph{Web and Internet Economics}.
  Cham, Switzerland: Springer International Publishing, pp. 106--119.

\bibitem[{Blunden(2016)}]{Blunden2016}
Blunden A (2016) \emph{The Origins of Collective Decision-Making},
  \emph{Studies in critical social sciences}, volume~84.
\newblock Leiden, Netherlands: Brill Academic Publishing.

\bibitem[{B{\"o}ttcher and Kernell(2022)}]{bottcher2022examining}
B{\"o}ttcher L and Kernell G (2022) Examining the limits of the {C}ondorcet
  {J}ury {T}heorem: Tradeoffs in hierarchical information aggregation systems.
\newblock \emph{Collective Intelligence} 1: 26339137221133401.

\bibitem[{B\"ottcher and Klingebiel(2024)}]{boettcher2024selection}
B\"ottcher L and Klingebiel R (2024) The organizational selection of
  innovation.
\newblock \emph{Organization Science} .

\bibitem[{Brams and Taylor(1996)}]{Brams1996}
Brams SJ and Taylor AD (1996) \emph{{Fair division: From cake-cutting to
  dispute resolution}}.
\newblock Cambridge, UK: Cambridge University Press.

\bibitem[{Brandt et~al.(2016)Brandt, Conitzer, Endriss, Lang and
  Procaccia}]{Brandt2016}
Brandt F, Conitzer V, Endriss U, Lang J and Procaccia AD (eds.)  (2016)
  \emph{Handbook of Computational Social Choice}.
\newblock Cambridge, UK: Cambridge University Press.

\bibitem[{Cacchiani et~al.(2022)Cacchiani, Iori, Locatelli and
  Martello}]{Cacchiani2022}
Cacchiani V, Iori M, Locatelli A and Martello S (2022) {Knapsack problems. {An}
  overview of recent advances. Part I: Single knapsack problems}.
\newblock \emph{Computers $\&$ Operations Research} 143: 105692.

\bibitem[{Csaszar and Eggers(2013)}]{csaszar2013organizational}
Csaszar FA and Eggers JP (2013) Organizational decision making: An information
  aggregation view.
\newblock \emph{Management Science} 59: 2257--2277.

\bibitem[{Dantzig(1957)}]{dantzig1957discrete}
Dantzig GB (1957) Discrete-variable extremum problems.
\newblock \emph{Operations Research} 5: 266--277.

\bibitem[{Dudzi{\'n}ski and Walukiewicz(1987)}]{dudzinski1987exact}
Dudzi{\'n}ski K and Walukiewicz S (1987) Exact methods for the knapsack problem
  and its generalizations.
\newblock \emph{European Journal of Operational Research} 28: 3--21.

\bibitem[{Fluschnik et~al.(2019)Fluschnik, Skowron, Triphaus and
  Wilker}]{Fluschnik2019}
Fluschnik T, Skowron P, Triphaus M and Wilker K (2019) Fair knapsack.
\newblock \emph{Proceedings of the {AAAI Conference on Artificial
  Intelligence}} 33: 1941--1948.

\bibitem[{Gersbach(2005)}]{gersbach2005designing}
Gersbach H (2005) \emph{Designing Democracy: Ideas for Better Rules}.
\newblock Berlin, Heidelberg, Germany: Springer.

\bibitem[{Gersbach(2017)}]{gersbach2017redesigning}
Gersbach H (2017) \emph{Redesigning Democracy: More Ideas for Better Rules}.
\newblock Cham, Switzerland: Springer.

\bibitem[{Gersbach(2024)}]{gersbach2024forms}
Gersbach H (2024) Forms of new democracy.
\newblock \emph{Social Choice and Welfare} 62(4): 799--837.

\bibitem[{Glass(1979)}]{Glass1979}
Glass JJ (1979) Citizen participation in planning: {T}he relationship between
  objectives and techniques.
\newblock \emph{Journal of the American Planning Association} 45: 180--189.

\bibitem[{Goel et~al.(2019)Goel, Krishnaswamy, Sakshuwong and
  Aitamurto}]{Goel2019}
Goel A, Krishnaswamy AK, Sakshuwong S and Aitamurto T (2019) Knapsack voting
  for participatory budgeting.
\newblock \emph{ACM Transactions on Economics and Computation} 7: 1--27.

\bibitem[{Hersh(1999)}]{hersh1999sustainable}
Hersh MA (1999) Sustainable decision making: the role of decision support
  systems.
\newblock \emph{IEEE Transactions on Systems, Man, and Cybernetics, Part C
  (Applications and Reviews)} 29: 395--408.

\bibitem[{Hotelling(1929)}]{hotelling1929stability}
Hotelling H (1929) Stability in competition.
\newblock \emph{The Economic Journal} 39: 41--57.

\bibitem[{Innes and Booher(2004)}]{Innes2004}
Innes JE and Booher DE (2004) Reframing public participation: {S}trategies for
  the 21st century.
\newblock \emph{Planning Theory and Practice} 5: 419--436.

\bibitem[{Jaiswal(2012)}]{jaiswal2012military}
Jaiswal NK (2012) \emph{Military Operations Research: Quantitative Decision
  Making}, volume~5.
\newblock New York, NY, USA: Springer Science \& Business Media.

\bibitem[{Kurvers et~al.(2023)Kurvers, Nuzzolese, Russo, Barabucci, Herzog and
  Trianni}]{Kurvers2023}
Kurvers RHJM, Nuzzolese AG, Russo A, Barabucci G, Herzog SM and Trianni V
  (2023) Automating hybrid collective intelligence in open-ended medical
  diagnostics.
\newblock \emph{Proceedings of the National Academy of Sciences} 120:
  e2221473120.

\bibitem[{Lijphart(1977)}]{Lijphart1977}
Lijphart A (1977) \emph{Democracy in Plural Societies: {A} Comparative
  Exploration}.
\newblock New Haven, CT: Yale University Press.

\bibitem[{Martello and Toth(1987)}]{martello1987algorithms}
Martello S and Toth P (1987) Algorithms for knapsack problems.
\newblock \emph{North-Holland Mathematics Studies} 132: 213--257.

\bibitem[{Nasako and Murakami(2006)}]{nasako2006}
Nasako T and Murakami Y (2006) A high-density knapsack cryptosystem using
  combined trapdoor.
\newblock \emph{Japan Society for Industrial and Applied Mathematics} 16:
  519--605.

\bibitem[{Nikolaos(2003)}]{Zahariadis2003}
Nikolaos Z (2003) \emph{Ambiguity and Choice in Public Policy: {P}olitical
  Decision Making in Modern Democracies}.
\newblock American Governance and Public Policy. Washington, DC: Georgetown
  University Press.

\bibitem[{Novshek(1982)}]{NOVSHEK1982199}
Novshek W (1982) Equilibrium in simple spatial (or differentiated product)
  models.
\newblock In: Mas-Colell A (ed.) \emph{Noncooperative Approaches to the Theory
  of Perfect Competition}. Academic Press, pp. 199--212.

\bibitem[{Pisinger(2005)}]{pisinger2005hard}
Pisinger D (2005) Where are the hard knapsack problems?
\newblock \emph{Computers \& Operations Research} 32: 2271--2284.

\bibitem[{Pisinger and Toth(1998)}]{Pisinger1998}
Pisinger D and Toth P (1998) Knapsack problems.
\newblock In: Du DZ and Pardalos PM (eds.) \emph{Handbook of Combinatorial
  Optimization}, volume 1--3. Boston, MA: Springer, pp. 299--428.

\bibitem[{Sah and Stiglitz(1986)}]{Sah1986}
Sah RK and Stiglitz JE (1986) The architecture of economic systems:
  {H}ierarchies and polyarchies.
\newblock \emph{The American Economic Review} 76: 716--727.

\bibitem[{Sah and Stiglitz(1988)}]{sah1988committees}
Sah RK and Stiglitz JE (1988) Committees, hierarchies and polyarchies.
\newblock \emph{The Economic Journal} 98: 451--470.

\bibitem[{Sanoff(2006)}]{Sanoff2006}
Sanoff H (2006) Multiple views of participatory design.
\newblock \emph{ODTÜ Mimarlık Fakültesi Dergisi} 23: 131--143.

\bibitem[{Schofield(2002)}]{Schofield2002}
Schofield NJ (2002) Representative democracy as social choice.
\newblock \emph{Handbook of Social Choice and Welfare} 1: 425--455.

\bibitem[{Schwenk(1990)}]{Schwenk1990}
Schwenk G (1990) Conflict in organizational decision making: {A}n exploratory
  study of its effects in for-profit and not-for-profit organizations.
\newblock \emph{Management Science} 26: 436--448.

\bibitem[{Smith-Miles et~al.(2021)Smith-Miles, Christiansen and
  Munoz}]{smith2021revisiting}
Smith-Miles K, Christiansen J and Munoz MA (2021) Revisiting {Where are the
  hard knapsack problems?} via instance space analysis.
\newblock \emph{Computers \& Operations Research} 128: 105184.

\bibitem[{Srivastava et~al.(2022)Srivastava, Pradhan and
  Saini}]{Srivastava2022}
Srivastava G, Pradhan N and Saini Y (2022) Ensemble of deep neural networks
  based on {C}ondorcet’s jury theorem for screening {Covid-19} and pneumonia
  from radiograph images.
\newblock \emph{Computers in Biology and Medicine} 149: 105979.

\bibitem[{{Stanford Crowdsourced Democracy Team}(2024)}]{PBStanford2024}
{Stanford Crowdsourced Democracy Team} (2024) \url{https://pbstanford.org/}.
\newblock Last accessed June 30, 2024.

\bibitem[{Stuart and Ord(2010)}]{stuart2010kendall}
Stuart A and Ord K (2010) \emph{Kendall's advanced theory of statistics,
  distribution theory}, volume~1.
\newblock 6$^{\rm th}$ edition. London, UK: Hodder Education Publishers.

\bibitem[{Wahlstr{\"o}m(2001)}]{Wahlstrom2001}
Wahlstr{\"o}m B (2001) Societal rationality; towards an understanding of
  decision making processes in society.
\newblock In: \emph{Values in Decisions on Risk}. Stockholm, Sweden, pp.
  369--381.

\bibitem[{Wampler(2007)}]{Wampler2007}
Wampler B (2007) A guide to participatory budgeting.
\newblock In: Shah A (ed.) \emph{Participatory Budgeting}. Washington, DC: The
  World Bank, pp. 21--54.

\end{thebibliography}
\end{document}